\documentclass[%
 amsmath,amssymb,
 reprint,superscriptaddress,%
]{revtex4-1}

\usepackage{graphicx}
\graphicspath{{Figures/}}
\usepackage[colorlinks,allcolors=blue]{hyperref}
\usepackage{dcolumn}
\usepackage{amsmath,amssymb,bm,mathtools,upgreek}
\usepackage{lineno}

\usepackage[utf8]{inputenc}
\usepackage[T1]{fontenc}
\usepackage{etoolbox}
\makeatletter
\def\@email#1#2{%
 \endgroup
 \patchcmd{\titleblock@produce}
  {\frontmatter@RRAPformat}
  {\frontmatter@RRAPformat{\produce@RRAP{*#1\href{mailto:#2}{#2}}}\frontmatter@RRAPformat}
  {}{}
}%
\makeatother
\usepackage{newtxtext,newtxmath,pdfrender}

\begin{document}

\preprint{JVSTA}

\title[]{Probing trade-off between critical size and velocity in cold-spray: An atomistic simulation}
\author{Mahyar Ghasemi}
\author{Alireza Seifi}%
\affiliation{Department of Materials and Metallurgical Engineering, Amirkabir University of Technology, 15875-4413, 15916-34311, Tehran, Iran}%
\author{Movaffaq Kateb}
\email{movaffaq.kateb@chalmers.se}
\affiliation{Condensed Matter and Materials Theory Division, Department of Physics, \\Chalmers University of Technology, SE-412 96 Gothenburg, Sweden}%
\author{Jon Tomas Gudmundsson}
\affiliation{Science Institute, University of Iceland,
Dunhaga 3, IS-107 Reykjavik, Iceland}
\affiliation{Division of Space and Plasma Physics, School of Electrical Engineering and Computer Science, \\KTH Royal Institute of Technology, SE-100 44, Stockholm, Sweden}
\author{Pascal Brault}
\affiliation{GREMI UMR7344 CNRS, Universit{\'e} d'Orl{\'e}ans, BP6744, 45067 Orleans CEDEX 2, France}
\author{Pirooz Marashi}
\affiliation{Department of Materials and Metallurgical Engineering, Amirkabir University of Technology, 15875-4413, 15916-34311, Tehran, Iran}
\date{\today}

\begin{abstract}
The detailed mechanism of bonding in the cold spray process has remained elusive for both experimental and theoretical parties. Adiabatic shear instability and hydrodynamic plasticity models have been so far the most popular explanations. Here, using molecular dynamics simulation, we investigate their validity at the nanoscale. The present study has potential application for the fabrication of ultra-thin layers for the electronics industry. For this aim, we considered Ti nanoparticles of different diameters and Si substrates of different orientations. It is shown that very high spray velocities are required for a jet to be observed at the nanoscale. We propose a method for thermostating the substrate that enables utilizing high spray velocities. For the first time, we demonstrate an oscillatory behavior in both the normal and radial stress components within the substrate that can propagate into the particle. We have shown that neither the adiabatic shear instability model nor the hydrodynamic plasticity model can be ignored at the nanoscale. Besides, the formation of a low-resistance titanium silicide proper for electronic application is illustrated.
\end{abstract}

\maketitle
\section{Introduction}
Cold spray \citep{assadi2003,adaannyiak23:69} is a deposition method in which particles are accelerated towards a substrate by supersonic expansion of a gas stream. The term \emph{cold} refers to the fact that, unlike thermal spray \citep{herman00:17}, there is no phase change, especially melting, prior to the particle impact onto the substrate and thus the adhesion is solely dependent on the spray velocity ($V_{0}$) that produces strain rates on the order of 1/ns \citep{assadi2003} during the impact. Despite the growing interest in the real-time imaging of the single particle collision \citep{adaannyiak23:69,hassani2018a}, the temporal and spatial resolution required for understanding the detailed mechanism is experimentally beyond the reach. Thus our understating of the cold spray process is limited to theories based on the post-deposition characterization.

To verify hypotheses regarding the detailed cold spray process mechanism, simulation and modeling have been shown promising (cf.~Refs.~\cite{assadi2003, rahmati2014, fardan2021}). In particular, the time scale in molecular dynamics (MD) simulations is highly compatible with that of the cold spray i.e.\ $V_{0}$ translates to 1 -- 3 lattice parameters per ps and strain rates on the order of 100/ps can be achieved \cite{jami2019, jami2021}. Besides, there is a growing interest in the nanoparticle cold spray technique \citep{samuel2017,wang2017,an2020,park2020,watanabe23:226}. MD simulations have been verified to capture nanoscale phenomena owing to their atomistic resolution \citep{kateb2012, kateb2018, azadeh2019, kateb2020b}. 

Several MD simulation studies have been conducted on the cold-spray process (cf.\ Ref.~\citep{fardan2021} and Refs.~therein). Probably \citet{gao2007} were the first to demonstrate simulations of the cold spray process by MD. They  pointed out that the process involves partial melting behavior.  
In an another interesting approach, \citet{daneshian2014} used the low-cost Lennard-Jones force field in 2D and modified its cut-off to model brittle particle impact on a rigid substrate. Their simple model showed a fair agreement with the analytical approach and captured many critical facts including plastic deformation below a certain size. 

There have been efforts to study nanoparticle cold spray using MD simulations to achieve ultra-thin, yet uniform films, suitable for the electronics industry. It has been shown that deposition under an angle, namely a larger off-normal angle (30$^\circ$), increases the film uniformity \citep{joshi2018}. However, the quality of the film, as measured by residual stress, becomes inferior due reduction in the normal component of the spray velocity $V_{0}$ \citep{james2020}. Thus, one may need to increase the spray velocity $V_{0}$ beyond the current limits to reach optimum uniformity and stress.

The plastic deformation correlation with local temperature and their dependence on the spray velocity $V_{0}$ were demonstrated by \citet{jami2019} using MD simulations. They also reported minimal changes with the particle size (2 and 20~nm) except for the fact that size-dependent melting temperature may contribute to the process. Later,  they studied the TiO$_2$/Ti system using a relatively expensive force field \citep{jami2021}. 
They observed plastic deformation of brittle TiO$_2$ in agreement with the simple 2D model \citep{daneshian2014}. \citet{rahmati2020} studied a wide range of particle sizes (5 -- 40~nm) for the Cu/Cu system and demonstrated localization of the dislocation network inside the particle next to the substrate. The possibility of mesoscale simulation and polycrystalline Al/Al system using quasi-course-grained has also been demonstrated \citep{suresh2020}. However, they maintained their substrate temperature at 150~K to avoid stability issues.  

Many of the above-mentioned studies are set up as a proof of principle studies,  so they consider a similar composition for both the nanoparticle and the substrate \citep{gao2007, joshi2018, james2020, rahmati2020, pereira2021}. Such an assumption normally leads to non-realistic interface behaviors. As we showed earlier \citep{kateb2019, kateb2020, kateb2021}, for Cu/Cu deposition by various methods, film-substrate similarity might overestimate intermixing even with high accuracy force fields such as EAM \cite{goel2014}. Moreover, the energy barrier for re-crystallization of a pure amorphous interface is negligible which can exaggerate post-impact crystal recovery \citep{gao2007, pereira2021, goel2014}.

While some studies were justly focused on the effect of the spray velocity $V_{0}$, they delivered the results in a limited velocity window 300 -- 500~m/s, or even neglected the particle size effect (cf.~Ref.~\citep{ jami2021,james2020}). As pointed out by \citet{pereira2021}, a spray velocity $V_{0}$ over 500~m/s is suggested to break surface oxide/contamination of Cu particles. Earlier, \citet{assadi2003} proposed $V_{0}>$~550~m/s for \emph{adiabatic shear instability} to occur for Cu. The low $V_{0}$ has probably been a technical issue associated with the lack of, or improper substrate thermostating, as hinted by \citep{gao2007,rahmati2020}. In the method section we proposed utilizing a three-layer substrate scheme, a common method in high energy ion bombardment simulation, that allows performing cold spray at very high $V_{0}$. This is important because the drag force exerted by gas expansion is proportional to the particle's cross-section or surface area in general. However, the acceleration is inversely proportional to the particle's mass and thus to its volume. Assuming a spherical particle with diameter $D$ the acceleration becomes proportional to $1/D$. Therefore, smaller particles achieve higher ultimate velocities than larger ones \citep{assadi2003,gilmore1999,schmidt2006,moridi2014}.

In the present work, we study cold spray deposition of a Ti nanoparticle  onto a Si substrate. This is a very complex system due to the presence of metallic, covalent, and covalent-ionic bonds between TiTi, SiSi, and TiSi, respectively. Thus, one can expect a totally different chemistry than in the previous metal/metal system studies. Ti is a highly reactive metal that requires extra special considerations to be deposited by conventional thermal evaporation or sputtering. Thus, cold spray can be a unique alternative to minimize the oxide and nitride ratio. Rapid thermal annealing of Ti/Si results in reduced sheet resistance which is desirable for interconnects for the semiconductor industry. One can skip the annealing step if increased mixing is achieved through adjusting the spray velocity $V_{0}$. For these reasons, we believe Ti/Si cold spray is important from both scientific and engineering points of view. Here we focus on the single collision/deposition with an emphasis on the effect of $V_{0}$. Single particle collisions allow us to understand  the bonding mechanism and have become popular in practice \citep{adaannyiak23:69}. For the current study we consider different substrate orientations as well as different particle sizes.
The MD simulation method is described in Section \ref{sec:method},  the results are presented in Section \ref{sec:results}, and a summary is given in Section \ref{sec:summary}.

\section{Method} \label{sec:method}
Our MD simulations \citep{allen1989} were performed by Large-scale Atomic/Molecular Massively Parallel Simulator (LAMMPS) \citep{plimpton1995,plimpton2012} code \footnote{LAMMPS website, \url{http://lammps.sandia.gov/}, distribution 29-Oct-2020}. The 3-body Tersoff force field was employed to model the entire interactions of the system. The general form of \citet{tersoff1988} potential is described by:
\begin{equation}
    U(r_{ij})=f_{\rm c}(r_{ij})\big[A_{ij}\exp(-\lambda_{ij}r_{ij})-b_{ij}B_{ij}\exp(-\alpha_{ij}r_{ij})\big],
	\label{eq:short}
\end{equation}
where $r_{ij}$ is the distance between atom $i$ and $j$, and $A_{ij}$, $B_{ij}$, $\lambda_{ij}$ and $\alpha_{ij}$ are constant parameters and $f_{\rm c}$ is the smoothing function near the cut-offs. The $b_{ij}$ is the main bond order term of the Tersoff potential that affects the attraction upon the changes in the bond angle, number of nearest neighbors and their symmetry as follows:
\begin{equation}
    b_{ij}=\big[1+(\beta\zeta_{ij})^n\big]^{-\frac{1}{2n}}
\end{equation}
in which $\beta$ and $n$ are constants and $\zeta_{ij}$ is given by:
\begin{equation}
    \zeta_{ij}=\sum f_{\rm c}(r_{ij})g(\theta_{ijk})\exp\big[\lambda^m(r_{ij}-r_{ik})^m],
\end{equation}
where the exponential term considers how central atom $i$ energy due to atom $j$ changes with its 2nd nearest neighbor $k$ and thus $\lambda$ and $m$ are 3-body constants. The angle $\theta_{ijk}$ enters as:
\begin{equation}
    g(\theta_{ijk})=\left(1+\left(\frac{c}{d}\right)^2-\frac{c^2}{d^2+(h-\cos\theta_{ijk})^2}\right)\gamma_{ijk},
\end{equation}
where $c$, $d$, $h$ and $\gamma_{ijk}$, are constants. Note that some of these constants were only considered to adopt different existing formalisms and thus they are already known. We used the Tersoff parameters provided by \citet{plummer2019} which was originally developed for the Ti$_3$SiC$_2$ MAX phase. Such a structure consists of a sandwich of Si mono-layers between titanium carbide (Ti$_3$C$_2$) layers. In particular, the force field models pure Ti and Si very well and predicts their interface energy with unique precision, making it suitable for very high strain rates. The set of parameters compatible with LAMMPS can be found in the supplementary materials.

\begin{figure}[hbt!]
    \centering
    \includegraphics[width=.5\linewidth]{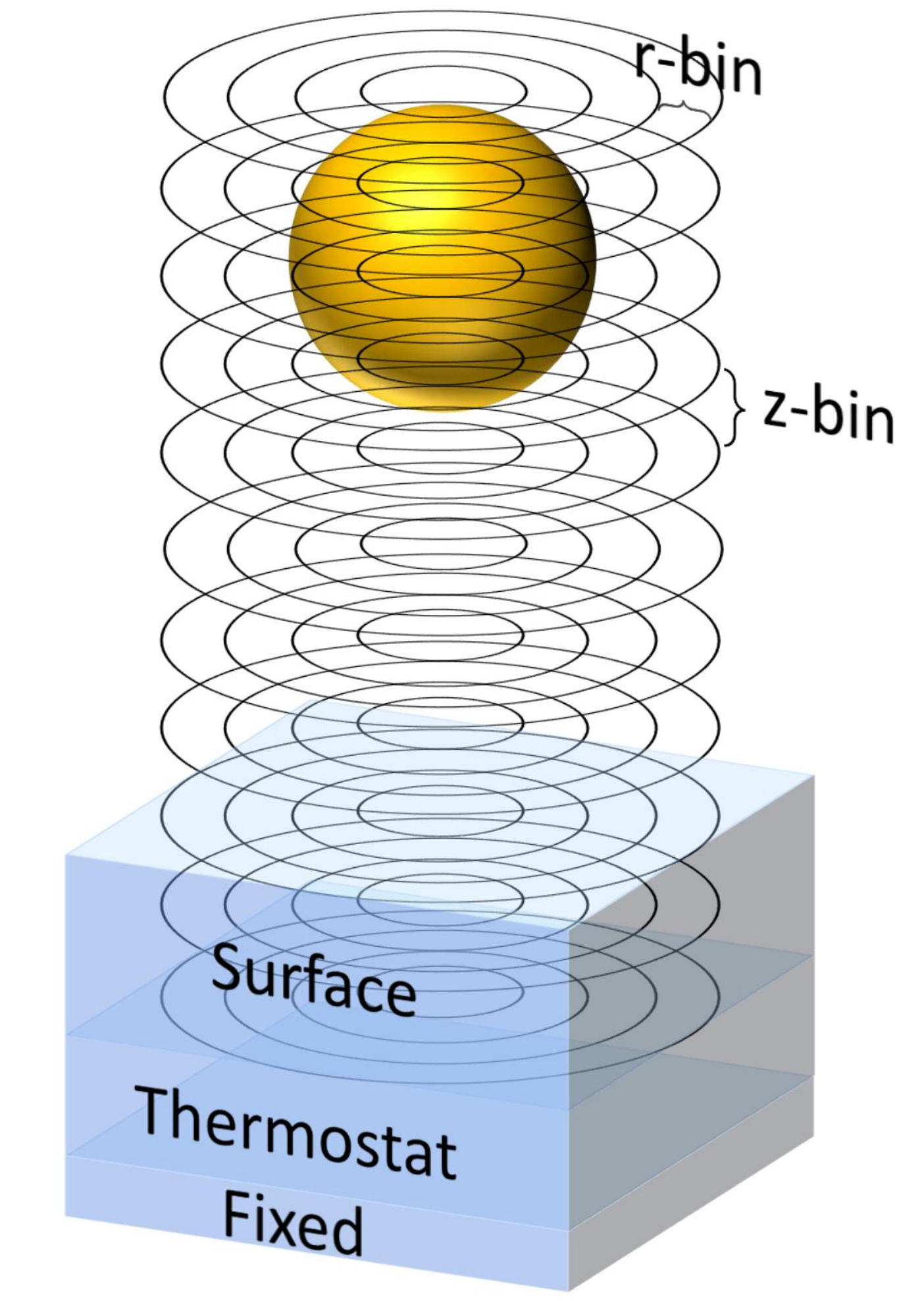}
    \caption{Schematic illustration of the cylindrical bins used for spatial averaging of the per-atom stress tensor $\boldsymbol{\upsigma}$. Note that $z$-bins span a range that includes both the particle and substrate surface and $r$-bins were considered from the center to the extent of the substrate plane.}
    \label{fig:bin}
\end{figure}

The substrate was assumed to be single crystal Si with nearly 100 $\times$ 100~{\AA}$^2$ lateral dimensions and 50~{\AA} thickness unless stated otherwise. We considered substrates with $\langle001\rangle$, $\langle011\rangle$, and $\langle111\rangle$ orientations along the growth direction ($z$).  Each substrate was allowed to expand in the plane ($xy$) to reach zero stress at 300~K before the cold spray experiment. Similarly, particles were relaxed at 300~K in a large enough box that resembles isolation in a vacuum. This is a sensible choice since nanoparticles have a lower melting point than bulk. Thus, to fully leverage the benefits of cold spray, it's crucial to prevent their melting. We divided each substrate into fixed, thermostat, and surface atoms as explained elsewhere \citep{kateb2019} and shown in Figure~\ref{fig:bin}. Briefly, the fixed layer is located at the bottom where we remove its net velocity but its atoms are allowed to vibrate naturally. The thermostat layer is meant to maintain the temperature of the surface layer at 300~K. The surface layer is thick enough so that particle collisions do not affect other layers. Such a scheme enables performing simulation at higher spray velocities $V_0$ than the earlier studies. 
Although many studies have omitted thermostatting, there are reports where it has been included \citep{gao2007,joshi2018,rahmati2020}.  \citet{gao2007} considered the fixed layer unnecessary since they studied very small particles. \citet{joshi2018,rahmati2020} considered the fixed layer, but treated it as completely static, without the natural atomic vibrations present at finite temperature. In the result section we explain how their assumptions kept them away from reaching to high spray velocities $V_0$ necessary for observation of the jet.

The Ti nanoparticles were assumed to have an icosahedral (ico) shape with diameters ranging from 0.9 to 4.1~nm. Each particle was individually relaxed at 300~K using the Nose-Hoover thermostat that generates samples from the canonical ensemble (NVT). Further details on the choice of shape an relaxation can be found elsewhere \citep{kateb2018, azadeh2019}.

For the time integration of the equation of motion, we follow the velocity Verlet algorithm \citep{kateb2012,verlet1967} using 0.5~fs timestep in both relaxations and the collision. The latter was performed by sampling from the microcanonical (NVE) ensemble. The thermostat layer was controlled by the Langevin thermostat \citep{schneider1978} with a damping of 1~fs and considering surface layer temperature. Such a small damping enables us to study very high spray velocity $V_0$ or up to 3000~m/s.

For quantitative comparison of intermixing, we calculated the partial radial distribution function, $g_{ij}(r)$ \citep{ashcroft1967}. However, the original notation is based on the binary mixtures being homogeneous.  For an earlier study we introduced a slightly modified version for an interface with continuous composition profile \citep{kateb2020} which will be used here.

The local microstructure is analyzed by polyhedral template matching (PTM) \citep{larsen2016}. Briefly, it generates a polyhedral shape using the 1st nearest neighbors and compares it to the desired template(s). It is insensitive to the interatomic distances used by other methods and thus immune to the error caused by strain and thermal fluctuation. However, similar to other methods, it distinguishes surface atoms as disordered, labeled as non, because of insufficient number of 1st nearest neighbors \citep{kateb2018,kateb2019}. We considered diamond cubic (dia), face-centered cubic (fcc), hexagonal closed packed (hcp), and ico templates. 

We utilized the open visual tool (Ovito) \citep{stukowski2009} code \footnote{Ovito website, \url{http://ovito.org/}, Version 3.0.0-dev794} to generate atomistic illustrations as well as for post-processing.

In the MD framework, per-atom stress tensor ($\boldsymbol{\upsigma}$) is calculated according to the \emph{virial} notation:

\begin{equation}
    \sigma_{pq} = -\frac{1}{\Omega_i}\left(m_iv_{p}v_{q}+\sum_{i\neq j}f_{ij}r_{ij}\right)
    \label{eq:virial}
\end{equation}
where $\sigma_{pq}$ is an element of $\boldsymbol{\upsigma}$ tensor with $p$ and $q$ being Cartesian coordinates, $m$ and $\Omega$ are atomic mass and volume of the reference atom $i$, respectively, $v_{p/q}$ is the velocity component of atom $i$, $f_{ij}$ is the force exerted by atom $j$ on atom $i$. The above definition can be understood as an outcome of the equation state for ideal gas plus excess pressure term due to interatomic forces. 

In order to convert the stress tensor $\boldsymbol{\upsigma}$ in Cartesian coordinates into cylindrical coordinates, we first must determine in-plane angle $\theta$. This can be simplified when the particle and substrate are centered at $x = y = 0$ and the particle moves along the $z$-axis as
\begin{equation}
    \theta=\arctan(y/x),
    \label{eq:theta}
\end{equation}
where $x$ and $y$ are the atom coordinates and thus an angle $\theta$ will be assigned to each atom. Now one can determine per-atom stress in cylindrical coordinates:
\begin{equation}
    \begin{aligned}
        \sigma_{rr} &= \cos^2{\theta}\sigma_{xx}+ 2\cos{\theta}\sin{\theta}\sigma_{xy} + \sin^2{\theta}\sigma_{yy} \\
        \sigma_{r\theta} &= \cos{\theta}\sin{\theta}(\sigma_{yy}-\sigma_{xx}) + (\cos^2{\theta}- \sin^2{\theta})\sigma_{xy} \\
        \sigma_{rz} &= \cos{\theta}\sigma_{xz} + \sin{\theta}\sigma_{yz} \\
        \sigma_{\theta\theta} &= \sin^2{\theta}\sigma_{xx} - 2\cos{\theta}\sin{\theta}\sigma_{xy} + \cos^2{\theta}\sigma_{yy} \\
        \sigma_{\theta z} &= -\sin{\theta}\sigma_{xz} + \cos{\theta}\sigma_{yz} \\
        \sigma_{zz} &= \sigma_{zz}
    \end{aligned}
    \label{eq:convert}
\end{equation}
where similar and disimilar subscript denote normal and shear components of the stress tensor $\boldsymbol{\upsigma}$. Regardless of the coordinate system, interpreting per-atom stress tensors is not an easy task.  To provide a simpler view of the stress distribution we averaged per-atom values obtained by Eq.~(\ref{eq:convert}) in cylindrical bins schematically as shown in Figure~\ref{fig:bin}. We considered 5 radial bins (rings) of $\sim$1~nm width, labeled by the average $r$ as $\bar{r}=$~0.5, 1.6, 2.7, 3.7, and 4.8~nm, and 130 bins along the $z$-axis (0.1~nm each). We have tested several bin-sizes along the $z$-axis and it seems one may sacrifice important features above 0.5~nm. Within each bin, the $\boldsymbol{\upsigma}$ tensor is averaged over all atoms and time-averaged for 100~timestep. 

\section{Results} 
\label{sec:results}
\subsection{Caloric response}
\begin{figure}[hbt!]
    \centering  \includegraphics[width=1\linewidth]{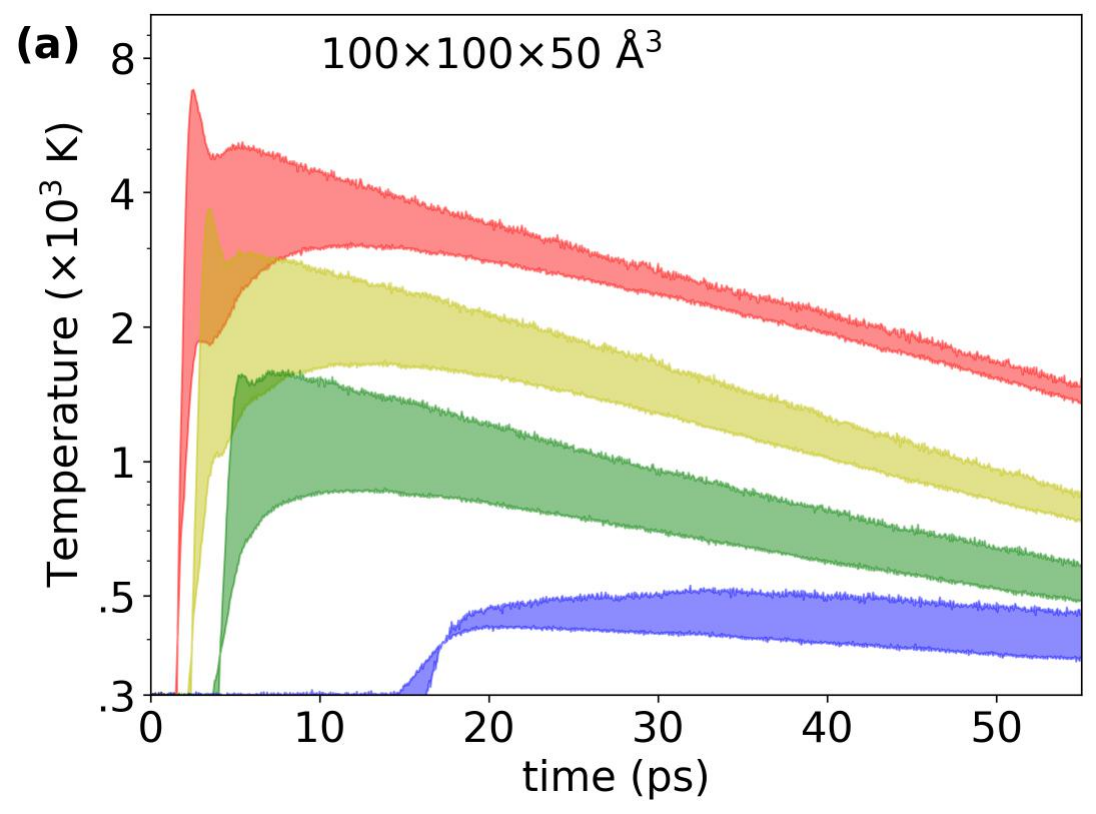}
    \includegraphics[width=1\linewidth]{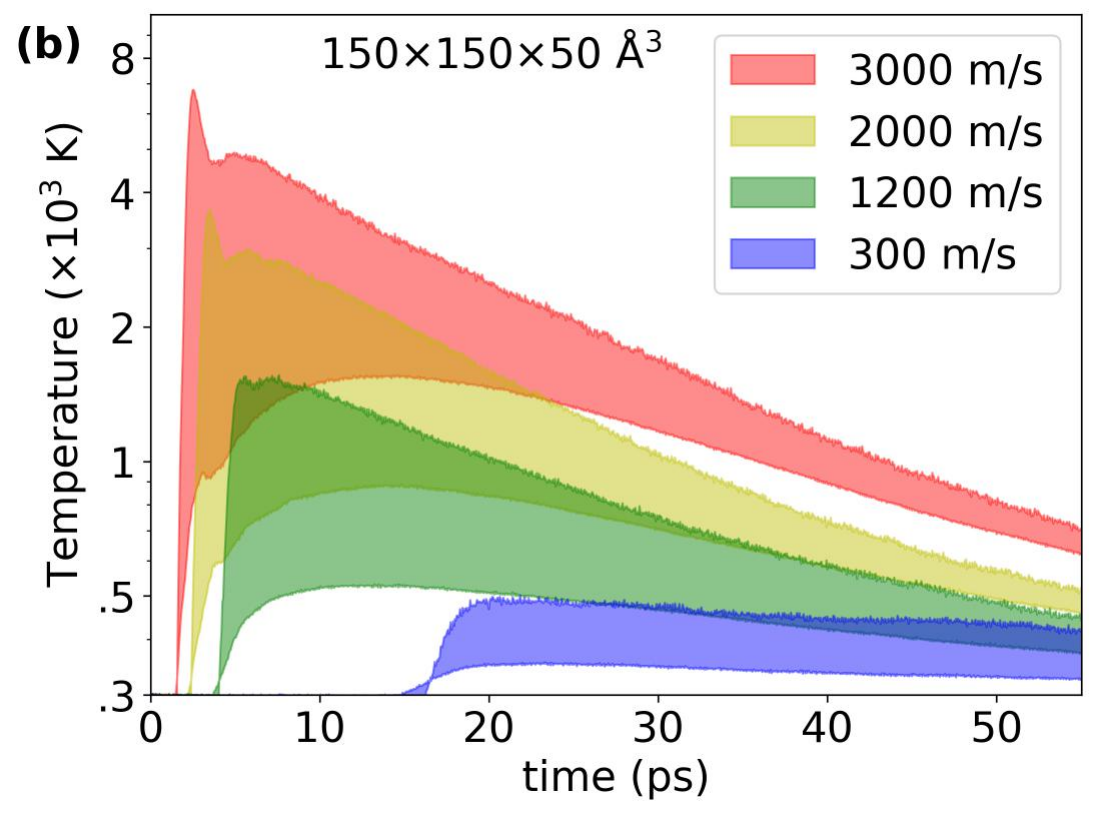}
    \caption{The temporal variation of $T_{\rm par}$ (top limit) and $T_{\rm sur}$ (bottom limit) with varying $V_{0}$ for a 4.1~nm diameter Ti ico particle onto a (100)~Si substrate with (a) 100 $\times$ 100, and (b) 150 $\times$ 150~\AA$^2$ lateral dimensions. Note that the vertical axis is plotted in the semilog.}
    \label{fig:tempico4.1}
\end{figure}

Figure~\ref{fig:tempico4.1} shows the temporal variation of temperature for various spray velocities $V_{0}$ for a 4.1~nm diameter Ti ico particle onto (100)~Si substrate. The top and bottom limits denote particle temperature ($T_{\rm par}$) and surface layer temperature ($T_{\rm sur}$), respectively. This defines the shaded area denoting the temperature difference between $T_{\rm par}$ and $T_{\rm sur}$. Since the surface layer is in contact with the thermostat (cf.~Figure~\ref{fig:bin}) its temperature is much better controlled and it stands a few hundred K below that of the particle. Figure~\ref{fig:tempico4.1}(a) shows the result for a substrate with 100$\times$100~\AA$^2$ lateral dimension in which we first observed a jet for $V_0$ of 3000~m/s. In order to make sure the jet is not an artifact of limited heat dissipation we repeated the numerical experiment with a larger substrate. For the larger substrate, shown in Figure~\ref{fig:tempico4.1}(b), the contact between the surface layer and the thermostat is larger, and consequently $T_{\rm sur}$ is even lower. However, the substrate dimensions do not change the early-stage variation of the particle temperature $T_{\rm par}$. At this stage, which will be referred to as impact hereafter, $T_{\rm par}$ is only determined by the spray velocity $V_{0}$. Upon the impact, two peaks are observed in the $T_{\rm par}$ at the higher velocities 1200 -- 3000~m/s. There is also a visible change in $T_{\rm sur}$ corresponding to the first peak in the $T_{\rm par}$.  
The range of temperatures observed here is in agreement with the findings of \citet{colla2000} but well below what was observed by \citet{cleveland1992}.

\begin{figure}[hbt!]
    \centering
    \includegraphics[width=1\linewidth]{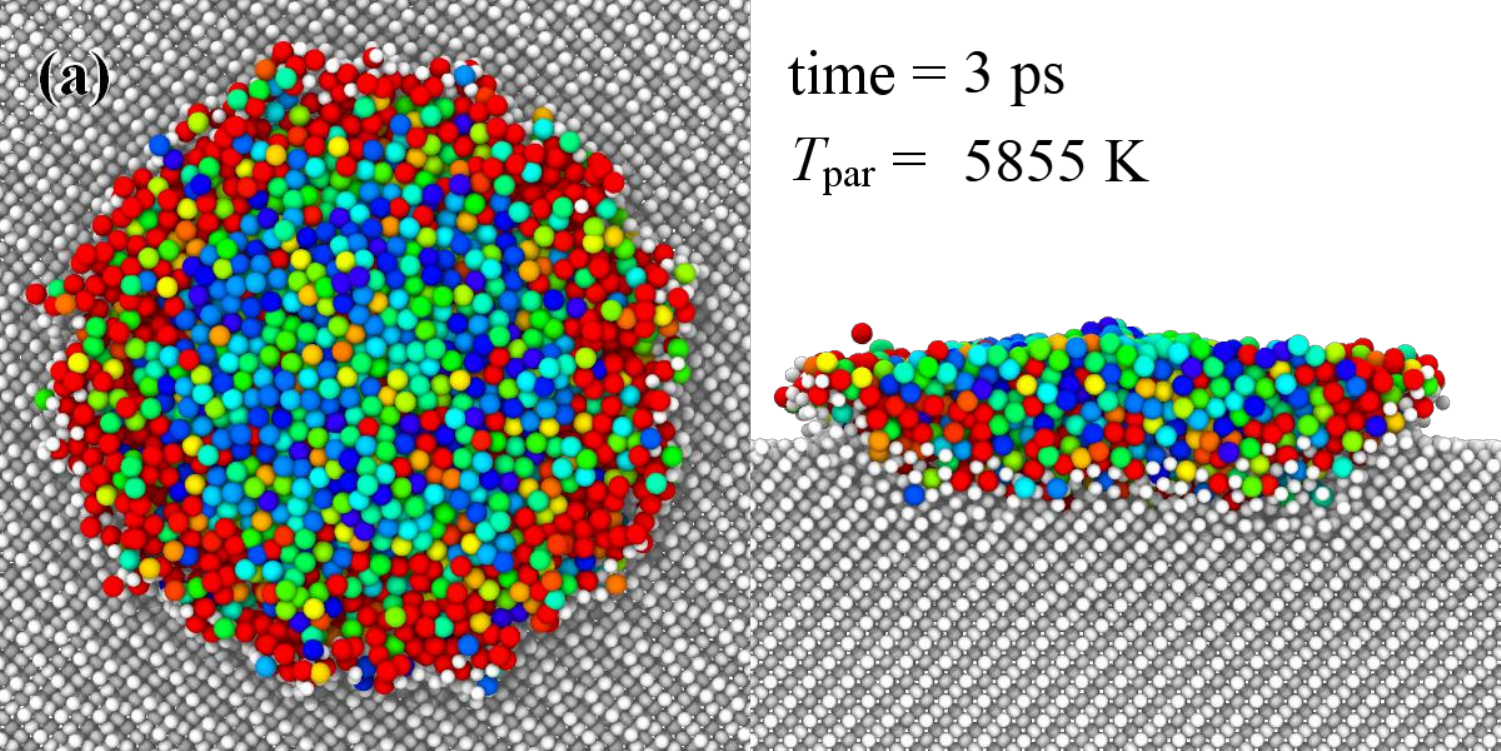}
    \includegraphics[width=1\linewidth]{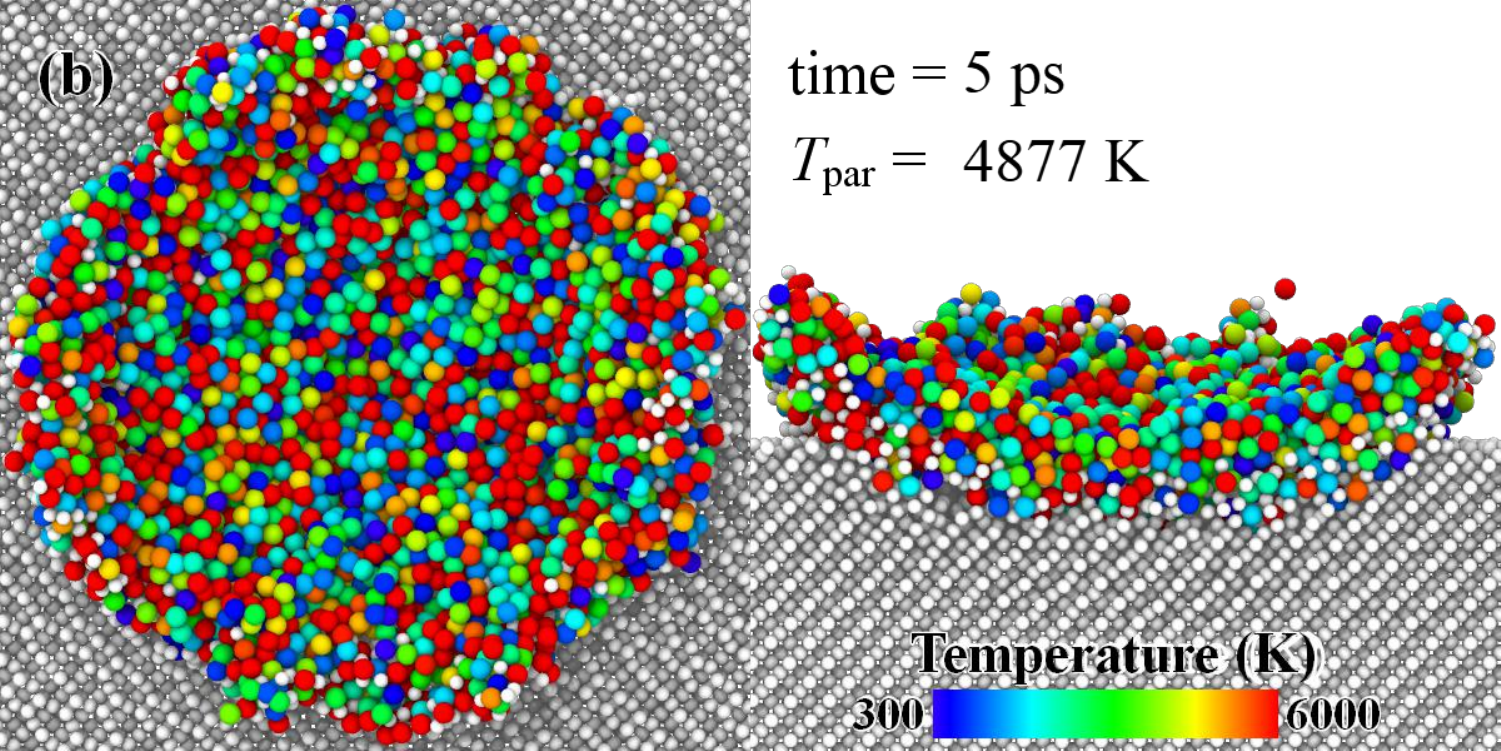}
    \caption{Atomistic illustrations of 4.1~nm diameter Ti ico particle onto a (100)~Si substrate with $V_{0}=$~3000~m/s. Particle top and side views for (a) flattening and (b) jet corresponding to the 1st and 2nd $T_{\rm par}$ peaks, respectively. For the side view, we made a slice through the middle of the particle. Substrate atoms are shown in gray for illustration purposes.}
    \label{fig:tempshots}
\end{figure}

The real-time comparison with the MD trajectory (Figure~\ref{fig:tempshots}) indicates that the 1st and 2nd $T_{\rm par}$ peaks are associated with particle \emph{flattening} and \emph{jet}, respectively. The fact that $T_{\rm par}$ does not change with better heat dissipation (larger substrate) and observation of a jet are solid pieces of evidence for adiabatic shear instability  \citep{assadi2003}. We did not observe the jet for spray velocities below 3000~m/s.
As previously mentioned, some studies have found thermostatting essential to achieve high spray velocities $V_0$ \citep{gao2007,joshi2018,rahmati2020}. However, due to slightly different conditions in the fixed layer and thermostat damping, we were able to reach the spray velocity $V_0$ necessary for jet ejection.

In the post-collision regime, $T_{\rm par}$ decreases monotonically, which is a characteristic of heat conduction into the surface layer. Similarly, the $T_{\rm sur}$ depends on the heat removed from the surface by the thermostat layer. A larger substrate area improves the heat dissipation from surface and lowers $T_{\rm sur}$. Consequently $T_{\rm par}$ drops faster, as seen in Figure~\ref{fig:tempico4.1}(b). On the other hand, $T_{\rm sur}$ still approaches its maximum ($T_{\rm sur}^{\rm max}$) after the jet. As $T_{\rm sur}$ without proper thermostatting is uninformative, earlier studies sufficed to report $T_{\rm par}$, cf.~\citep{jami2019,jami2021}, and $T_{\rm sur}$ has been barely reported \citep{gao2007}. For Au spray on Au \citet{gao2007} observed peak in $T_{\rm sur}$ with a constant delay after the collision, independent of $V_0$. As can be seen here the time between 1st $T_{\rm par}$ peak and $T_{\rm sur}^{\rm max}$ increases with an increase in $V_0$. Such a gradual increase in $T_{\rm sur}$ is associated with the ongoing intermixing, i.e. titanium silicide layer thickening. It is also worth mentioning that at spray velocity of 300~m/s we observed landing, i.e.\ without the plastic deformation. This is phenomenologically different than the impact at higher $V_{0}$. Thus, one cannot expect impact peaks in $T_{\rm par}$ at low $V_{0}$, as observed in earlier studies (cf.~\citep{jami2019,jami2021}).
We obtained a similar behavior for higher surface densities, i.e.~29\% and 13\% higher, corresponding to (110) and (111) Si substrates, respectively (cf. Figure S1 in supplementary materials).

\subsection{Bonding}
Figure~\ref{fig:diffuse} shows the dependence of inter-diffusion on the spray velocity $V_{0}$, for a 4.1~nm Ti diameter ico particle on (100), (110), and (111)~Si substrates. Considering the topmost Si atom in blue and the lowermost Ti atom in red, one can clearly see that the diffusion decreases with higher planar densities. This is even true at 3000~m/s where severe intermixing occurs.
\begin{figure}[hbt!]
    \centering
    \includegraphics[width=1\linewidth]{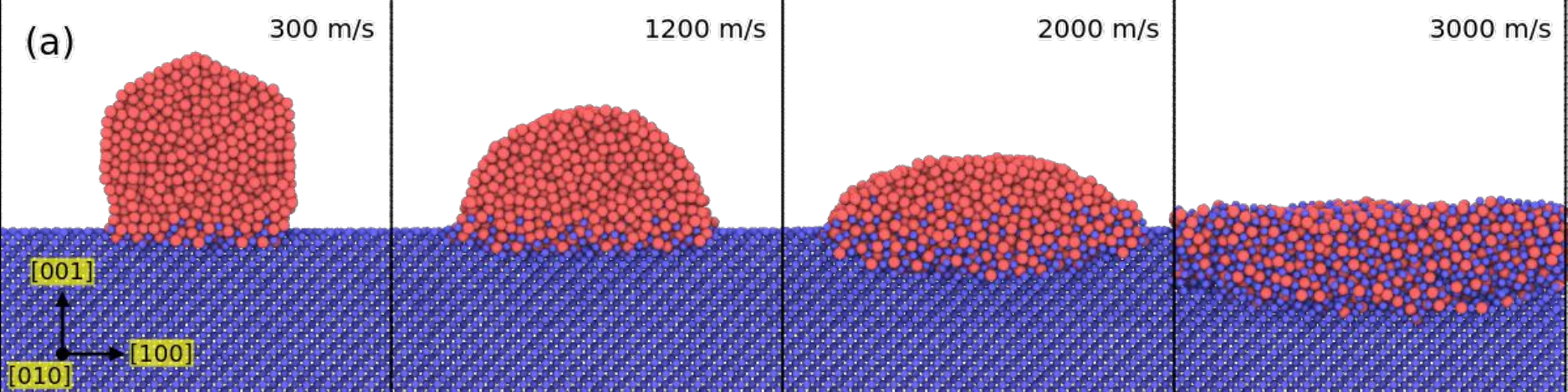}
    \includegraphics[width=1\linewidth]{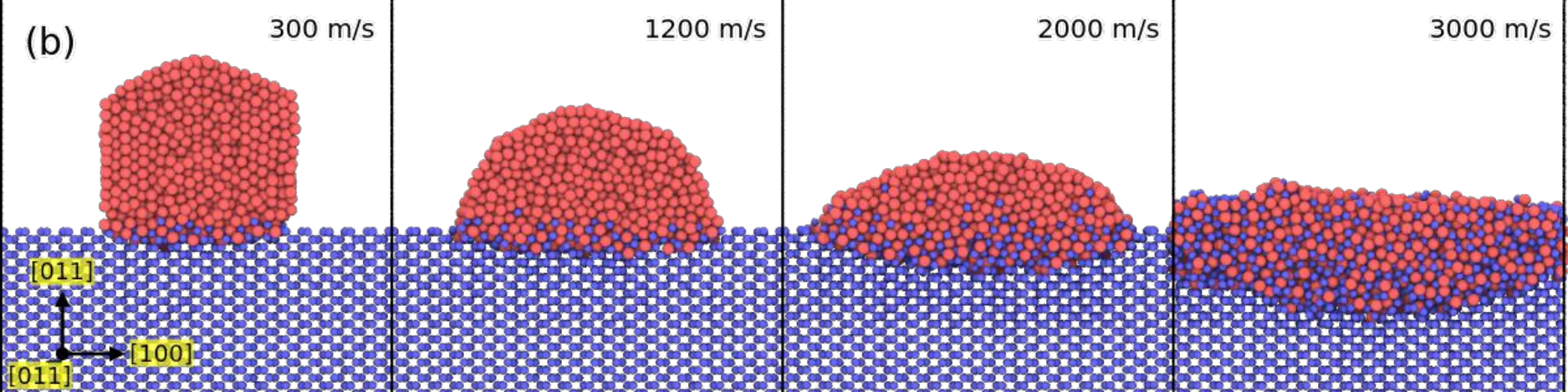}
    \includegraphics[width=1\linewidth]{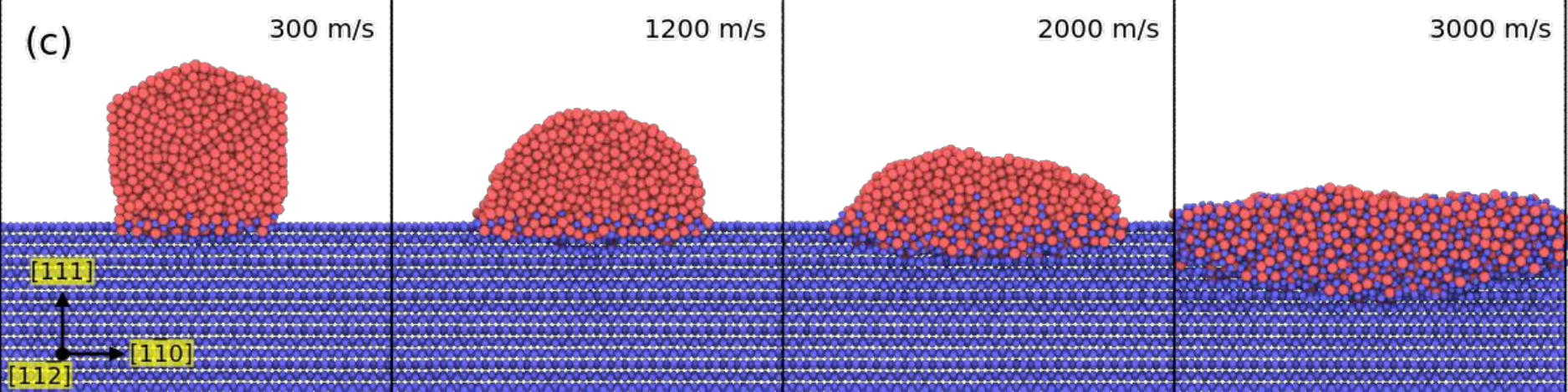}
    \caption{Dependence of inter-diffusion on the spray velocity $V_{0}$ for a 4.1~nm diameter  Ti ico particle on (a) (100), (b) (110), and (c) (111)~Si substrates. All snapshots depict a slice in the middle of particles.}
    \label{fig:diffuse}
\end{figure}

\begin{figure}[hbt!]
    \centering
    \includegraphics[width=1\linewidth]{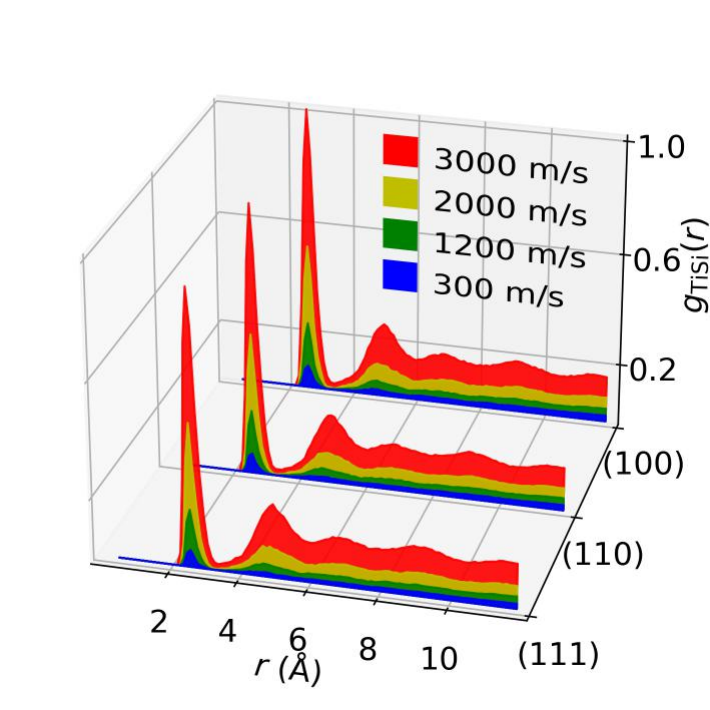}
    \caption{Partial radial distribution $g(r)$ for TiSi after spraying 4.1~nm diameter Ti ico particles onto (100), (110), and (111)~Si substrates.}
    \label{fig:gr4.1}
\end{figure}

While Figure~\ref{fig:diffuse} clearly depicts the depth of diffusion, complementary information on the inter-diffusion can be obtained by exploring the partial radial distribution, $g(r)$. Figure~\ref{fig:gr4.1} shows $g_{\rm TiSi}(r)$ for 4.1~nm diameter Ti ico particle onto different Si substrates. Looking at the major peak (at $r=$~2.5~{\AA}) one can see immediately that higher $V_{0}$ leads to more TiSi bonds (1st nearest neighbors). It can be seen that there is almost no difference between $g_{\rm TiSi}(r)$ obtained with different substrates. One may think this is inconsistent with the diffusion depth observed in Figure~\ref{fig:diffuse}. However, as explained elsewhere \citep{azadeh2019}, $g_{\rm TiSi}(r)$ is not sensitive to the detailed geometry. As a matter of fact, when the planar density is higher, a shorter diffusion depth will produce the same number of TiSi pairs. Thus, it is necessary to compare $g(r)$ with other information to acquire a full picture of the process. 

There have been efforts to predict critical velocity $V_{\rm cr}$ for bonding to occur based on the thermo-mechanical properties of the materials \citep{assadi2003, schmidt2006, hassani2018b}. \citet{schmidt2006} introduced, assuming a simple empirical model, $V_{\rm cr}=FD^{-n}$ at constant temperature, with $F$ and $n$ being fitting parameters and $D$ being particle diameter. Recently, \citet{dowding2020} fitted $F$ and $n$ for Ti particles of 5 -- 50~$\mu$m in diameter to be 1280~$\tfrac{\rm m/s}{\mu {\rm m}^{-n}}$ and 0.21, respectively. Extrapolating to 4.1~nm diameter particle size, a critical velocity of 4060~m/s can be estimated. This is in close agreement with the jet we observed at 3000~m/s considering the fact that we have ideally clean surfaces while experimental studies suffer from this point. We would like to remark that meeting the size and velocity criteria are necessary for adiabatic shear instability. Outside these choices, bonding may be achieved but with a different mechanism. For instance, although Au has a very large critical diameter, \citet{gao2007} observed successful bonding of its nanoparticles, attributing this to partial melting during the process.

\subsection{Spreading}
\begin{figure}[hbt!]
    \centering
    \includegraphics[width=1\linewidth]{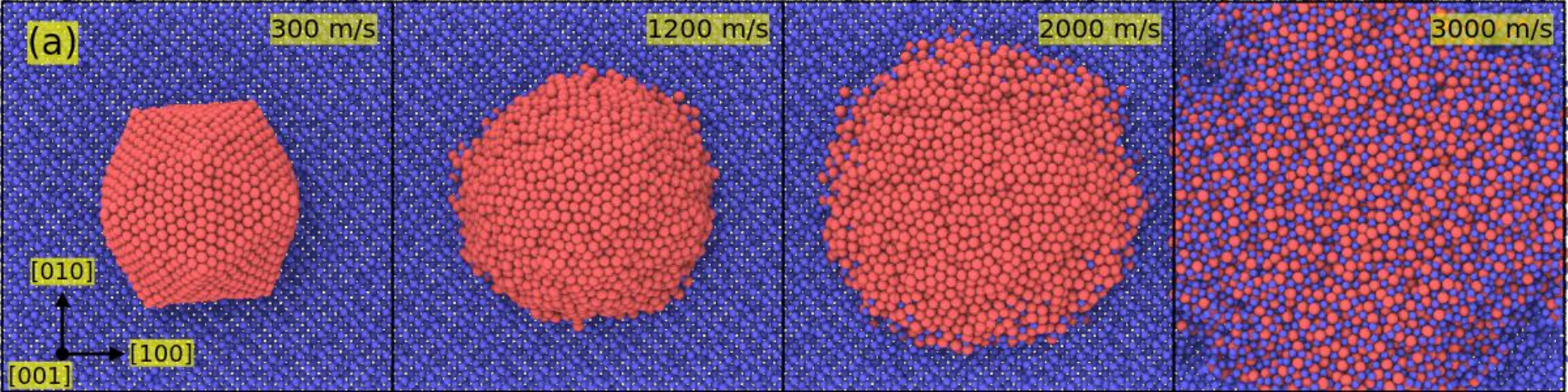}
    \includegraphics[width=1\linewidth]{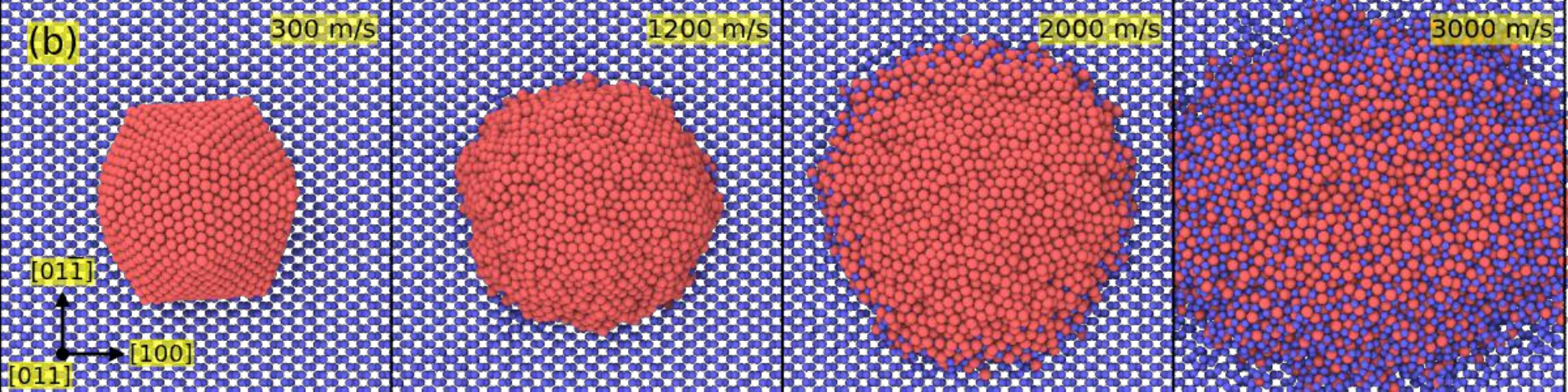}
    \includegraphics[width=1\linewidth]{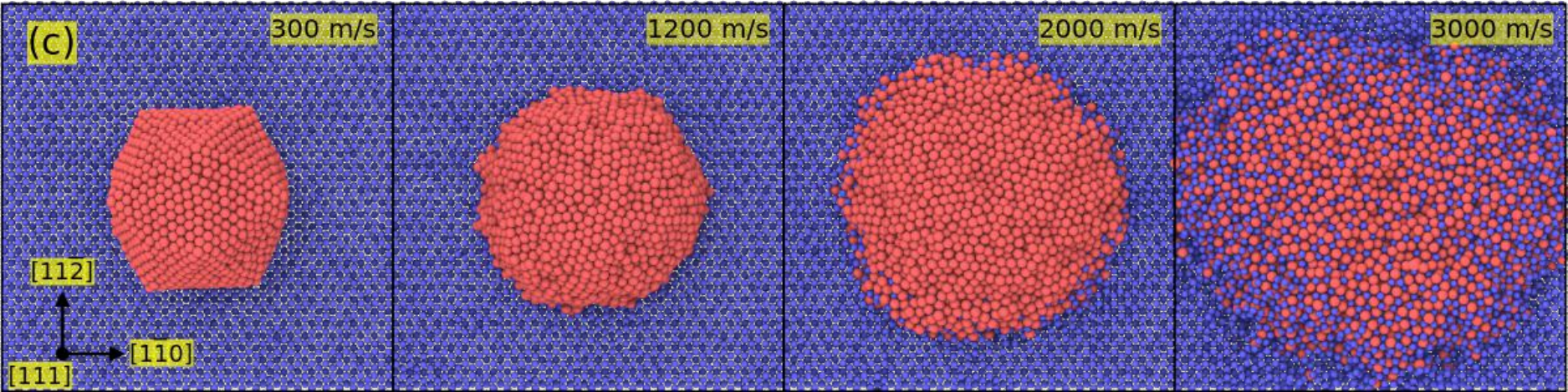}
    \caption{Dependence of the lateral spreading on the spray velocity $V_{0}$ for 4.1~nm diameter Ti ico particle spreading atop (a) (100), (b) (110), and (c) (111)~Si substrates.}
    \label{fig:spread}
\end{figure}

In practice the initial particle diameter ($D$) is unknown, thus the measurement of lateral spreading ($w$) and height ($h$) can be used to determine $D=\sqrt[3]{w^2h}$ \citep{king2010}. In simulations, $D$ is known and thus the spreading ratio $w/D$ can be determined as a measure of strain. Figure~\ref{fig:spread} shows the variation of lateral spreading with $V_{0}$ for 4.1~nm Ti diameter ico particle over (100), (110), and (111)~Si substrates. By fitting a circle to the outermost Ti atoms, as explained in the supplementary materials, spreading ratios of 1.3, 1.6, 2.0 and 2.5 were obtained at 300 -- 3000~m/s, with negligible changes on different substrates. It is worth mentioning that the particle nearly keeps the ico shape at $V_0$ of 300~m/s. Considering the five-fold symmetric corners in ico, they are still observable at the particle's top for 1200~m/s. At 2000~m/s, the corners' symmetry becomes vague while at 3000~m/s mixing with the substrate takes over. This is consistent with the earlier experimental \citep{hassani2018a} and simulation \citep{rahmati2020} studies that limit the plastic deformation to the bottom side of the particle. 
The spreading ratio is a rough measure of the strain meaning it represents an average value of local strains. The fact that the particle's top experiences smaller deformation indicates that the local strain at the interface is much larger than the spreading ratio.

\subsection{Microstructure}
\begin{figure}[hbt!]
    \centering
    \includegraphics[width=.49\linewidth]{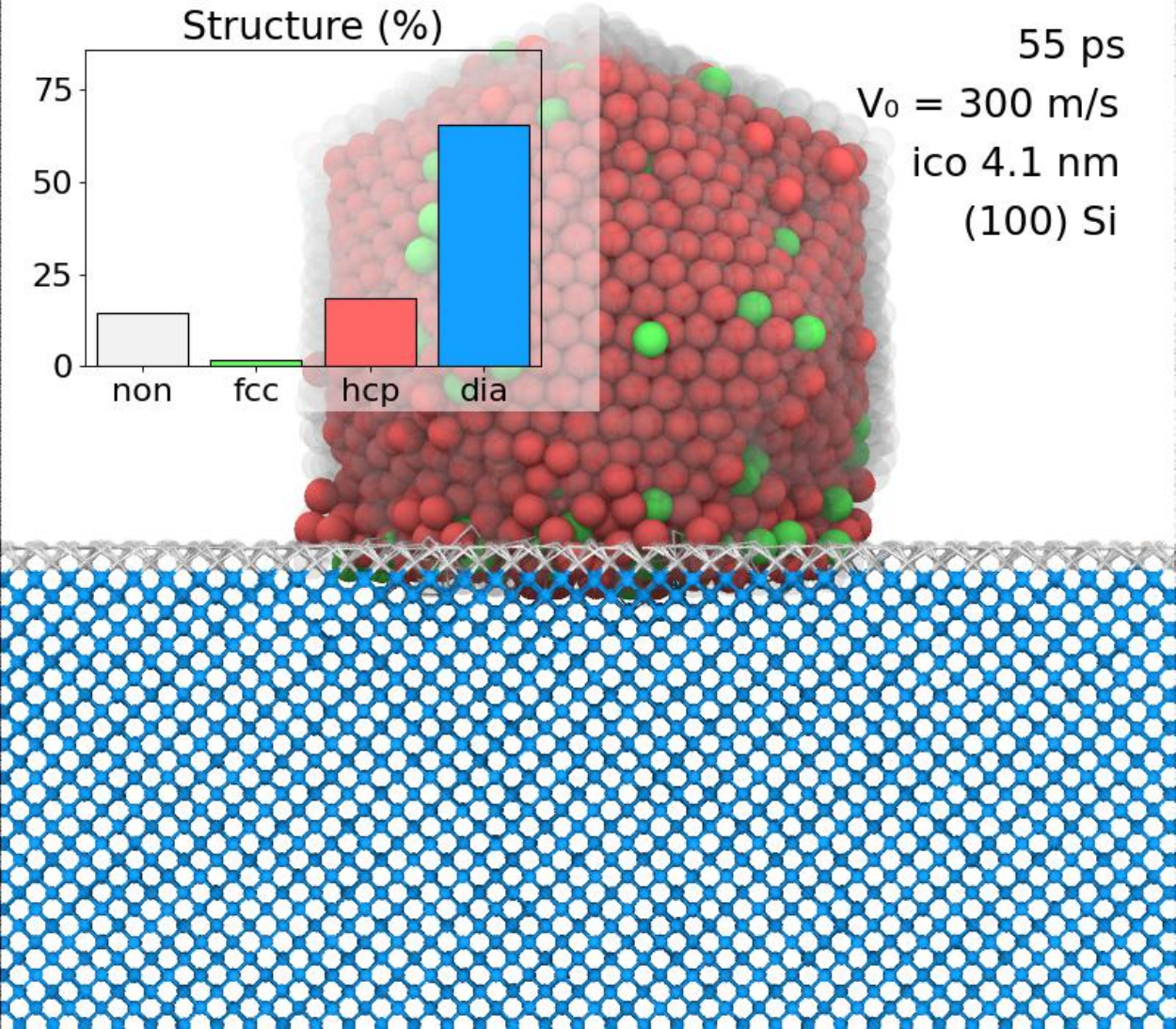}
    \includegraphics[width=.49\linewidth]{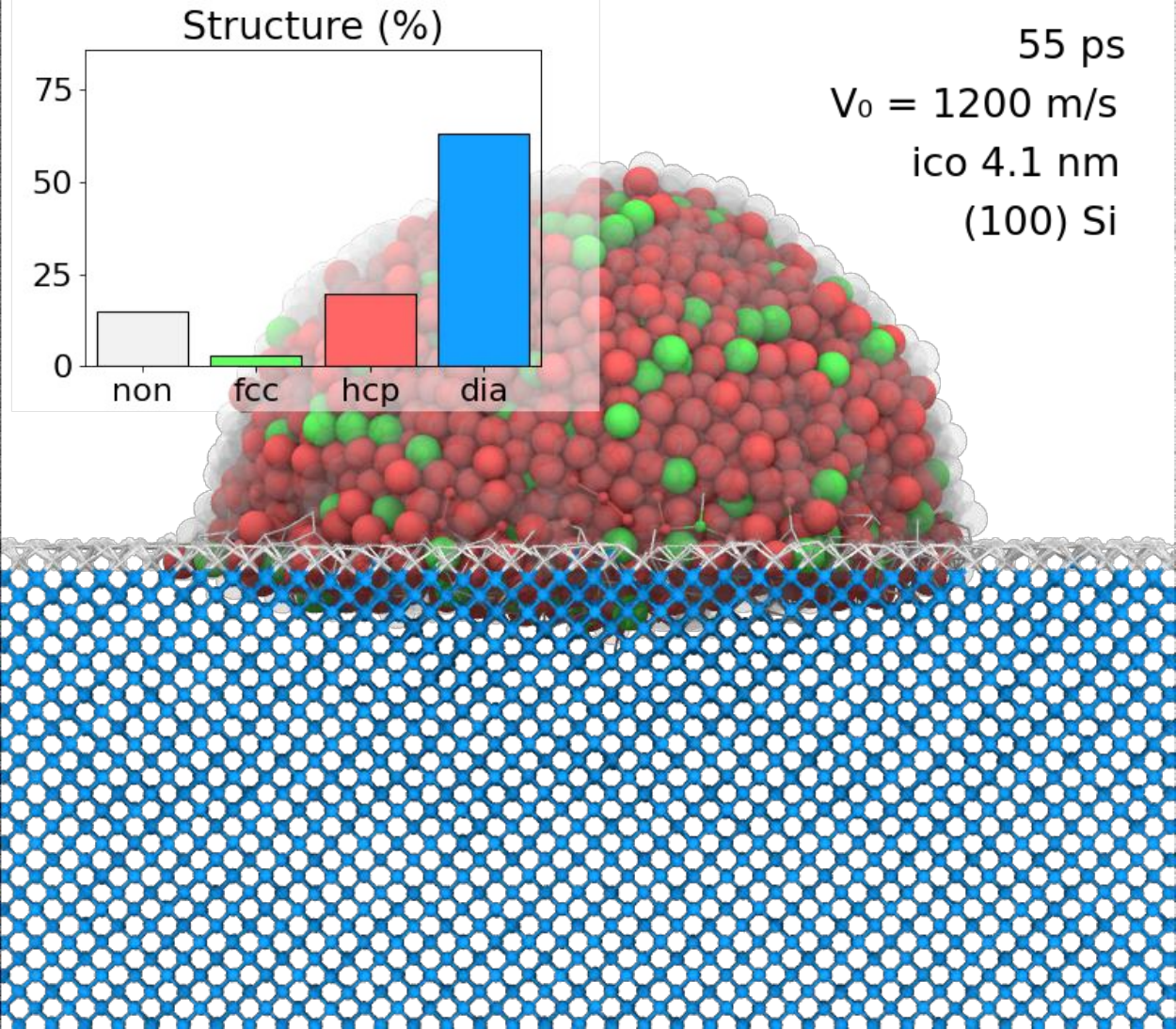}
    \includegraphics[width=.49\linewidth]{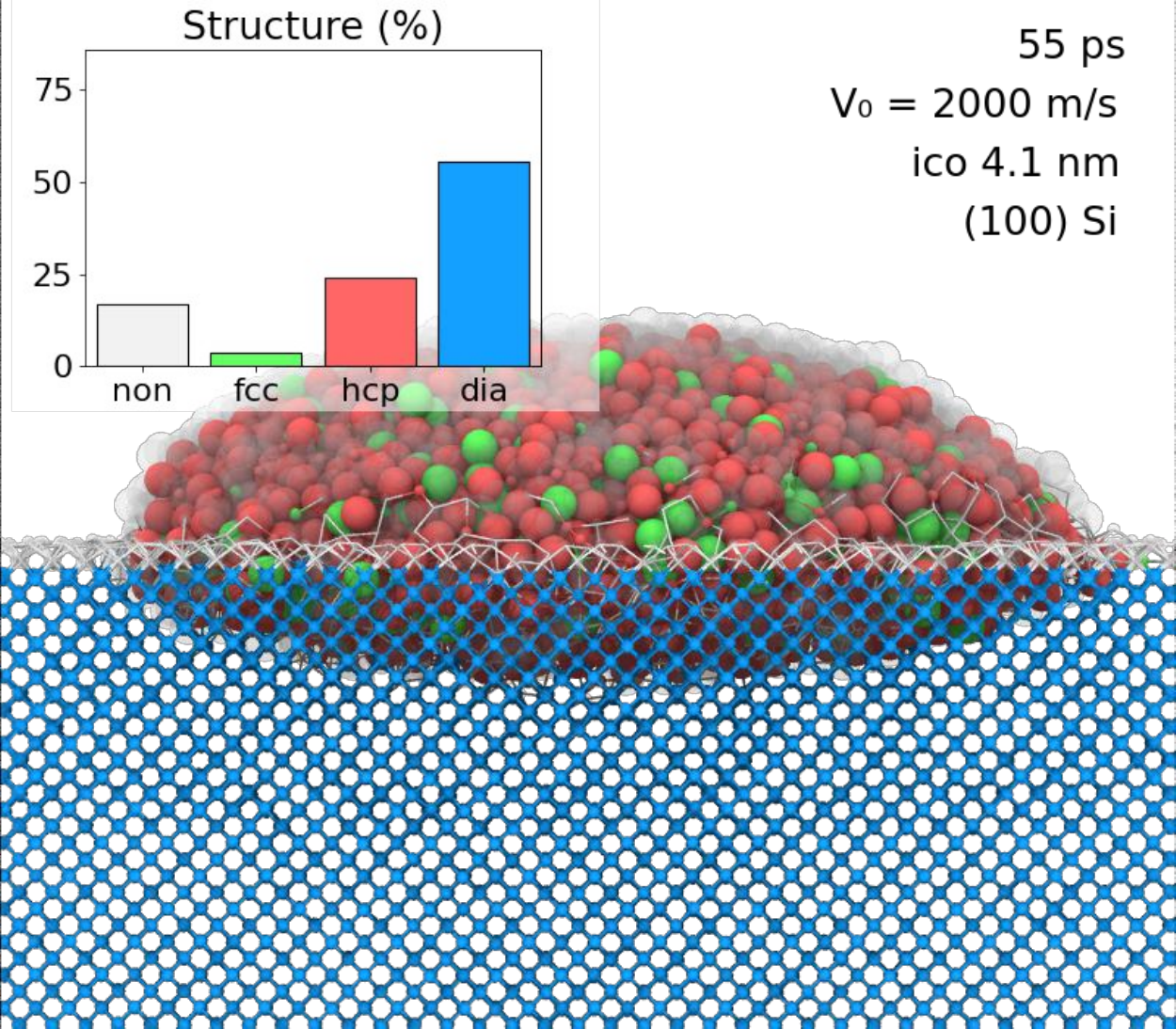}
    \includegraphics[width=.49\linewidth]{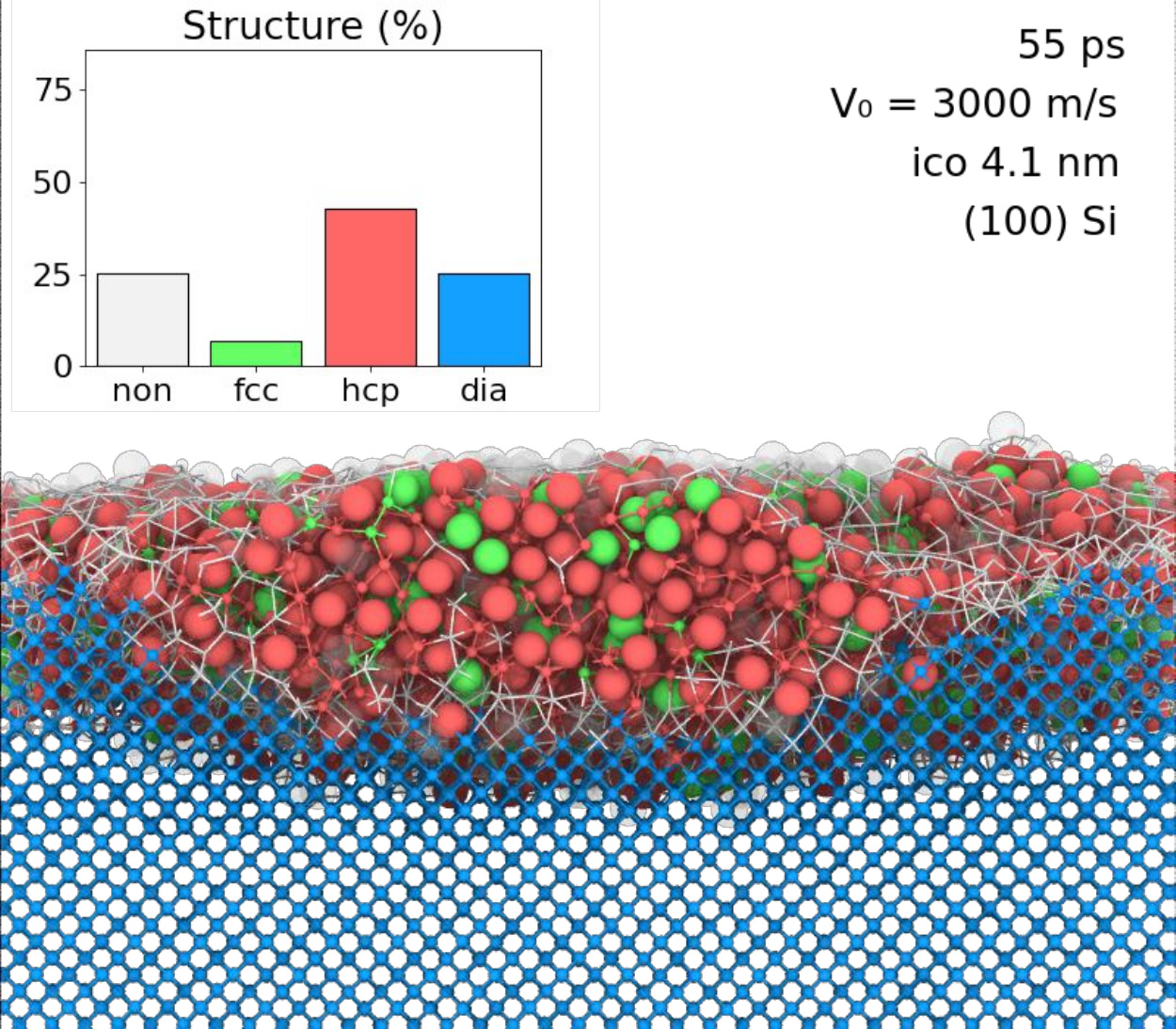}
    \caption{The result of PTM for 4.1~nm diameter Ti ico particle spray at different $V_{0}$ onto (100) Si substrate. Substrate atoms are shown with smaller diameters also surface and disordered atoms are presented as semi-transparent in white.}
    \label{fig:ptm}
\end{figure}

Figure~\ref{fig:ptm} shows a PTM analysis of 4.1~nm diameter Ti ico particles on a (100)~Si substrate, delivered at different spray velocities $V_{0}$. Note that we only included the particle and the surface layer in the PTM calculation. It can be seen that the ratio of non, fcc, and hcp structures increase with increased spray velocity $V_{0}$ at the cost of a reduction in the dia structure. At spray velocity of 3000~m/s the dia ratio reaches that of non and is well below that of hcp. The former can be expected in an energetic impact while the latter seems unpredicted. This can be explained considering the TiSi$_2$ structure as follows.

\begin{figure}[hbt!]
    \centering
    \includegraphics[width=.7\linewidth]{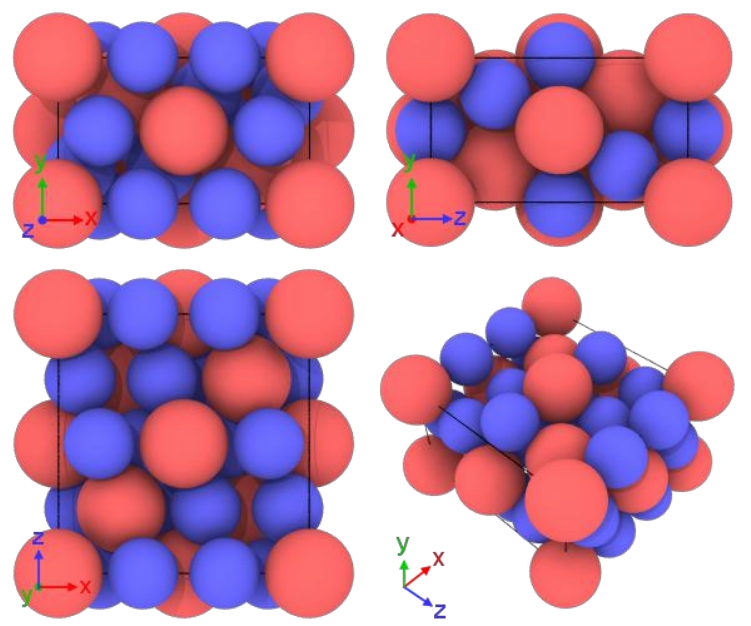}
    \caption{A C54 unit cell (8.2671 $\times$ 4.800 $ \times$ 8.5505~{\AA}$^3$) of TiSi$_2$ with Red and blue being Ti and Si, respectively. It is classified in the orthorhombic category with the oF24 Pearson symbol or Fddd space group. Note how the $xy$ plane (top-left) is compatible with the hcp basal plane.}
    \label{fig:C54}
\end{figure}

In the semiconductor industry, TiSi$_2$ is known to have C54 and C49 structures \citep{liao2006,mann93:3566}. The development of the high-resistivity (60 -- 90~${\upmu}\Omega$.cm) metastable C49 phase and its subsequent transformation into the low-resistivity (12 -- 20~$\upmu\Omega$.cm) C54 phase is of significant interest from both technological and scientific perspectives \citep{mann93:3566}. Here we emphasize on face-centered orthorhombic (C54) TiSi$_2$ as shown in Figure~\ref{fig:C54}. In the C54 structure, each Ti atom is surrounded by 10 Si atoms at 2.55 -- 2.78~{\AA} distance while Si atoms have 5 Ti and 5 Si nearest neighbors with Si-Si distance of 2.52 -- 2.79~{\AA}. The existence of 4 bond lengths makes the PTM characterization of TiSi$_2$ nearly impossible. However, it can be seen that the $xy$ plane (top-left) is highly compatible with the hcp basal plane. Thus, it is very likely that C54 is characterized as an hcp-like structure.

\subsection{Stress}
\begin{figure}[hbt!]
    \centering
    \includegraphics[width=1\linewidth]{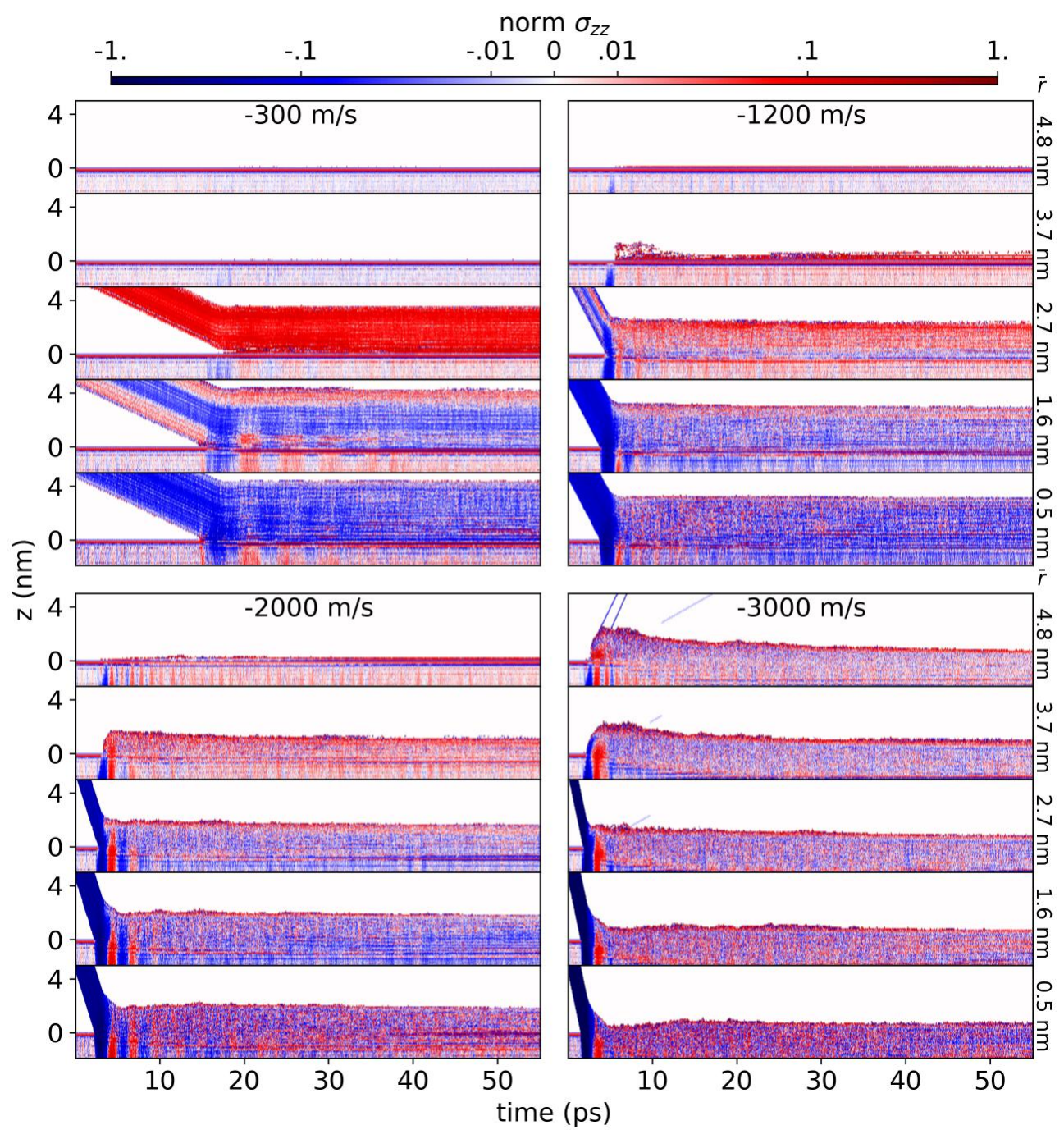}
    \caption{Variation of the normal stress component $\sigma_{zz}$ pattern by increase in $V_{0}$ for a 4.1~nm diameter ico particle and (001)~Si substrate. Note that the colorbar is in logscale outside $\pm$0.01.}
    \label{fig:s_zz}
\end{figure}

Figure~\ref{fig:s_zz} shows the variation of the normal stress component $\sigma_{zz}$ for 4.1~nm diameter Ti ico particle on (001)~Si substrate. The figure includes 4 panels, one for each given spray velocity $V_{0}$ (labeled on top), and within each panel, there are 5 subplots corresponding to the rings ($\bar{r}=$ 0.5, 1.6, 2.7, 3.7, and 4.8~nm, as indicated on the right side).  Here $z \leq 0$ corresponds to the substrate surface and the top branch ($z>0$) gradually approaching the substrate indicates the particle. Although both fcc and hcp structures have been reported for Ti nanoparticles \citep{xiong2010}, the hcp remains the most stable phase at 300~K. Here, we chose the ico shape, which predominantly exhibits a fcc structure, as it most closely resembles a spherical shape while still presenting similar corners and facets. However, ico shape produces compressive stress in the core and tensile stress at the particle's surface. Before the collision, at 300~m/s, the stress component $\sigma_{zz}$ at the particle's surface ($\bar{r}=2.7$~nm)  is tensile (in red), whereas in the particle's core it is compressive ($\bar{r}=$~0.5 and 1.6~nm). As $V_0$ increases, the stress component $\sigma_{zz}$ shifts towards a more compressive regime.

As expected, a particle impact introduces a compressive pulse into the substrate. The impact pulse appears as blue ($z<0$) when the particle meets the substrate and its intensity and duration depend on the spray velocity $V_{0}$. For instance, at 300~m/s the impact pulse appears very weak (pale blue) at $\bar{r}=$~3.7~nm while at higher $V_{0}$, it can propagate to the outermost ring. Besides, the duration of the impact pulse is shorter for  higher $V_{0}$ but more compressive (darker). 
It can be seen that almost always the normal stress component $\sigma_{zz}$ shows an oscillatory behavior, i.e.\ the impact pulse is followed by a red one and so on, those are clearly depicted for spray velocity of 2000~m/s at different $\bar{r}$. 
Earlier, \citet{hassani2018b} had argued against the adiabatic shear instability mechanism of bonding, originally proposed by \citet{assadi2003}. Instead, they suggested hydrodynamic plasticity at the interface that is induced by pressure waves. This is still an ongoing debate (cf.~Ref.s \citep{assadi2019,hassani2019}), but here for the first time, we demonstrate pressure waves which we refer to as stress oscillations. 
We believe the impact pulse produces a longitudinal (with respect to the substrate normal) acoustic wave for the following reasons: First, it is mostly incoherent at the interface ($z=0$). Second, it lasts longer on the substrate than the particle due to the well-defined lattice. Last but not of least importance, it decays faster with higher residual stress (darker red and blue dots).

The blue lines in the outer rings, observed for spray velocity of  3000~m/s, indicate jet ejection. Regardless of the detailed mechanism \citep{assadi2003,hassani2018b}, it appears that the jet has been ruled out as the necessary characteristics of bonding \citep{assadi2019,hassani2019}. We observed bonding for the different $V_{0}$ but only observed the jet at 3000~m/s in agreement with above-mentioned studies. Here we did not observe the rebound but we hypothesize the long-lasting normal stress component $\sigma_{zz}$ oscillations to be responsible for the rebound.

We would like to remark that, while earlier studies consider an overall stress picture (cf.~Ref.~\cite{jami2019, jami2021,james2020}) here we focused on the localized aspect that has been barely discussed \citep{assadi2003,daneshian2021}. For instance, at spray velocity of 300~m/s the substrate and the particle present different normal stress component $\sigma_{zz}$, above and below $z=0$. For higher $V_{0}$ we observe a more uniform normal stress component $\sigma_{zz}$ along the growth direction ($z$). Regardless of the velocity $V_{0}$, however, the normal stress component $\sigma_{zz}$ is more compressive in the inner rings and \textit{vice versa}. 
As pointed out earlier, an issue with the same particle and substrate is neglecting the interface effects. Here, it can be clearly seen that when moving along the $z$-direction, the normal stress component $\sigma_{zz}$ becomes discontinuous at the particle-substrate interface. The only exception is that  at spray velocity of 3000~m/s, where severe intermixing occurs,~i.e.\ we were unable to locate a cluster of pure Ti.

\begin{figure}[hbt!]
    \centering
    \includegraphics[width=1\linewidth]{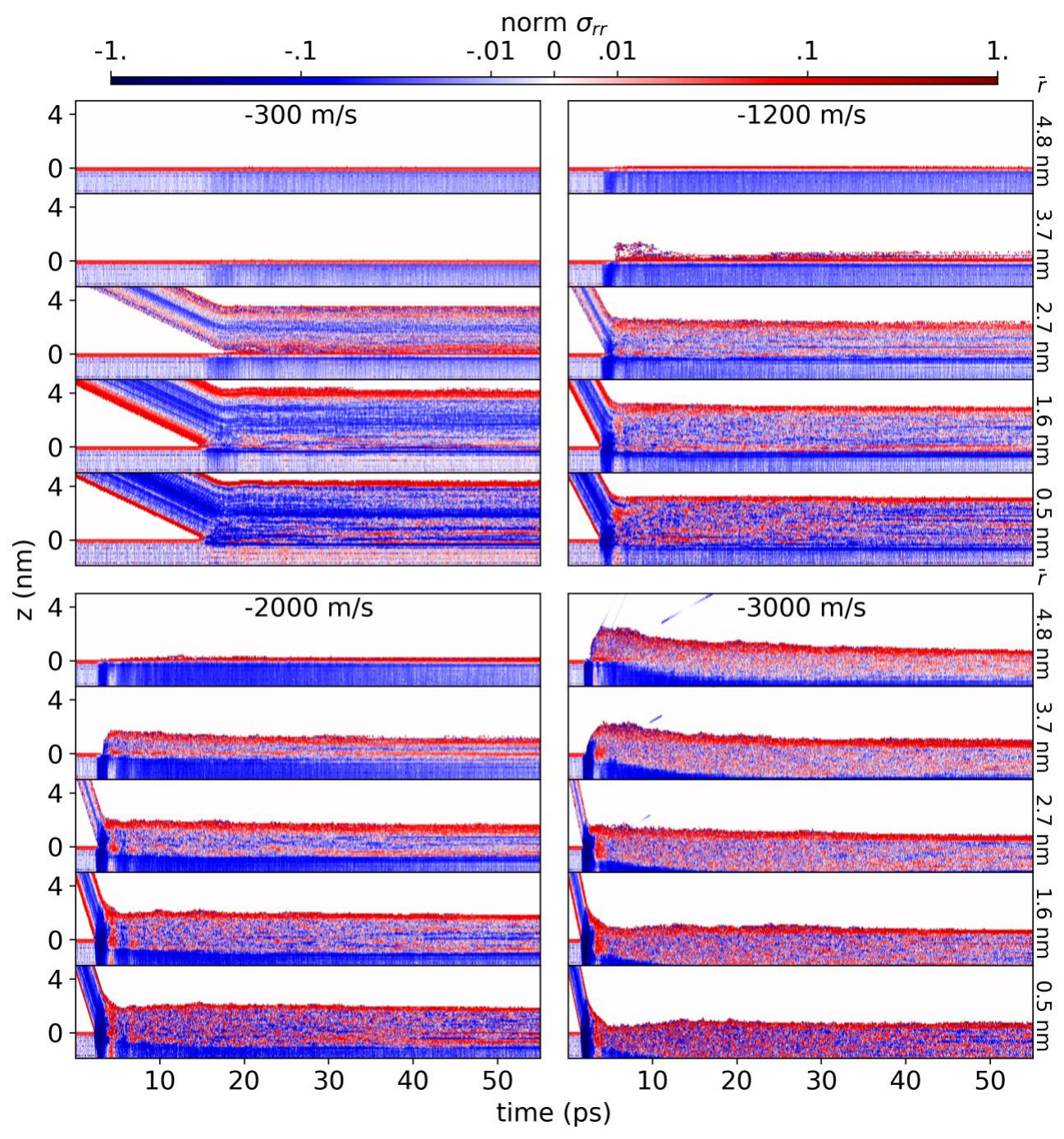}
    \caption{Variation of the radial stress component $\sigma_{rr}$ pattern by increase in $V_{0}$ for a 4.1~nm diameter Ti ico particle onto a (001)~Si substrate. Note that the colorbar is in logscale outside $\pm$0.01.}
    \label{fig:s_rr}
\end{figure}

The variation of the radial stress component $\sigma_{rr}$ at different values of $V_{0}$ is depicted in Figure~\ref{fig:s_rr} for a 4.1~nm diameter ico particle on (001)~Si substrate. Before the collision, regardless of $V_{0}$ and $\bar{r}$, the radial stress component $\sigma_{rr}$ is always tensile (red) at the particle surface and compressive in the core. The topmost layer ($z_{\rm max}$) remains tensile even after the collision. Unlike for the normal stress component $\sigma_{zz}$ case, collision-induced oscillations in the radial stress component $\sigma_{rr}$ are barely symmetric and overall compressive. In the particle and interface, however, the radial stress component $\sigma_{rr}$ presents a mixed state both tensile and compressive (red and blue). Similar to the normal stress component $\sigma_{zz}$, by probing along the $z$-direction, the radial stress component $\sigma_{rr}$ shows a discontinuity at the interface. However, the effect of $V_{0}$ on the radial stress component $\sigma_{rr}$ is different than on the normal stress component $\sigma_{zz}$. Here, higher $V_{0}$ shift the substrate's radial stress component $\sigma_{rr}$ towards more compression while its effect on the particle is the opposite. Thus as $V_{0}$ increases, the slope of the radial stress component $\sigma_{rr}$ at the interface increases.

\subsection{Size-dependence}

\begin{figure}[hbt!]
    \centering
    \includegraphics[width=0.32\linewidth]{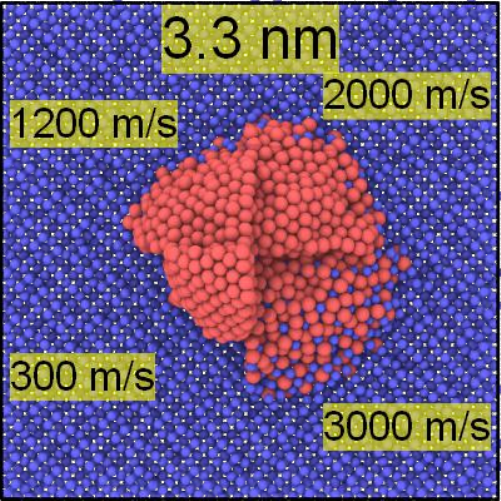}
    \includegraphics[width=0.32\linewidth]{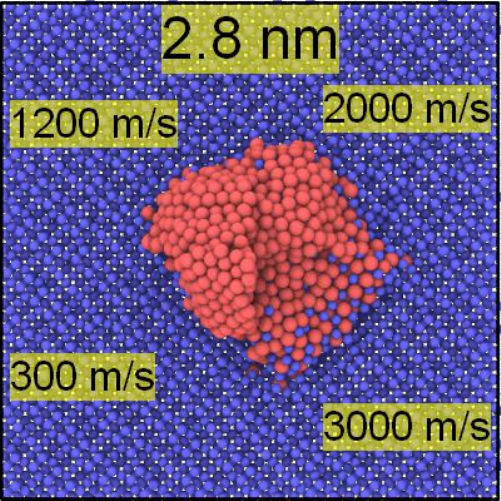}
    \includegraphics[width=0.32\linewidth]{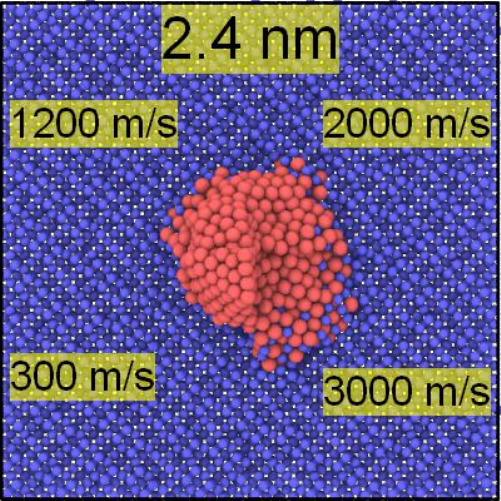}
    \includegraphics[width=0.32\linewidth]{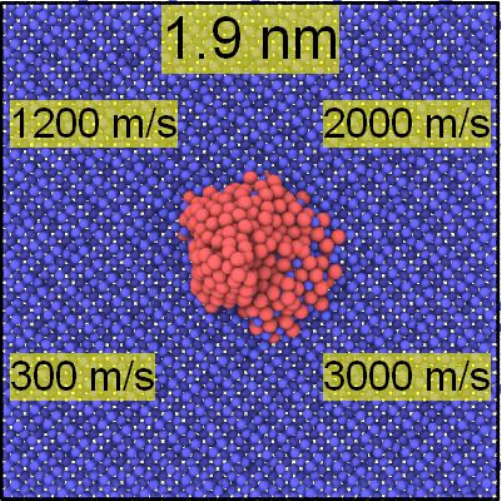}
    \includegraphics[width=0.32\linewidth]{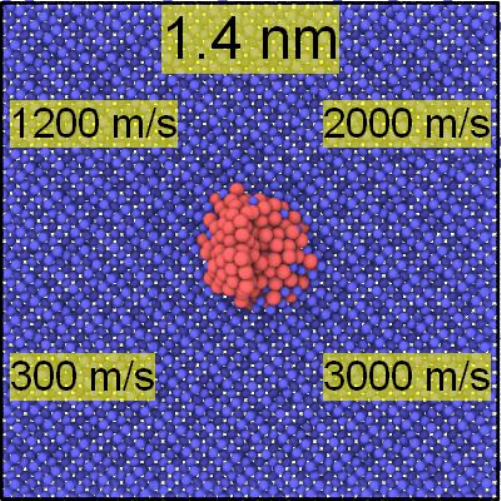}
    \includegraphics[width=0.32\linewidth]{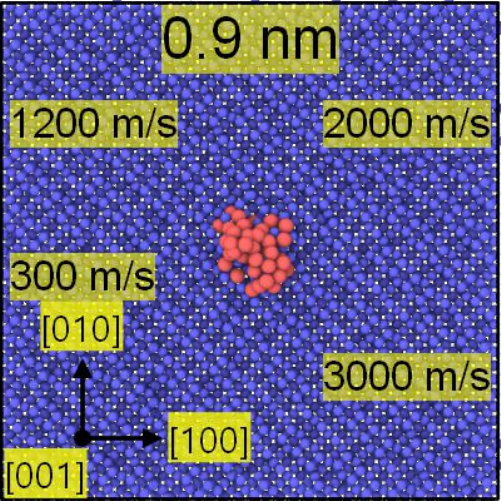}
    \caption{The dependence of lateral spreading on the size of the Ti ico particle. For each particle different spray velocities $V_{0}$ are shown in a quadrant.}
    \label{fig:spread_size}
\end{figure}

Figure~\ref{fig:spread_size} shows the variation of lateral spreading with the diameter of the Ti ico particle and the spray velocity $V_{0}$. For each particle size, denoted on top, four different spray velocities $V_0$ are shown as a quadrant. It can be seen that higher velocities are really ineffective in increasing the lateral spreading for the smallest particle, here 0.9~nm in diameter. However, for larger particles, an increase in the lateral spreading with higher $V_{0}$ is evident.

\begin{figure}[hbt!]
    \centering
    \includegraphics[width=.7\linewidth]{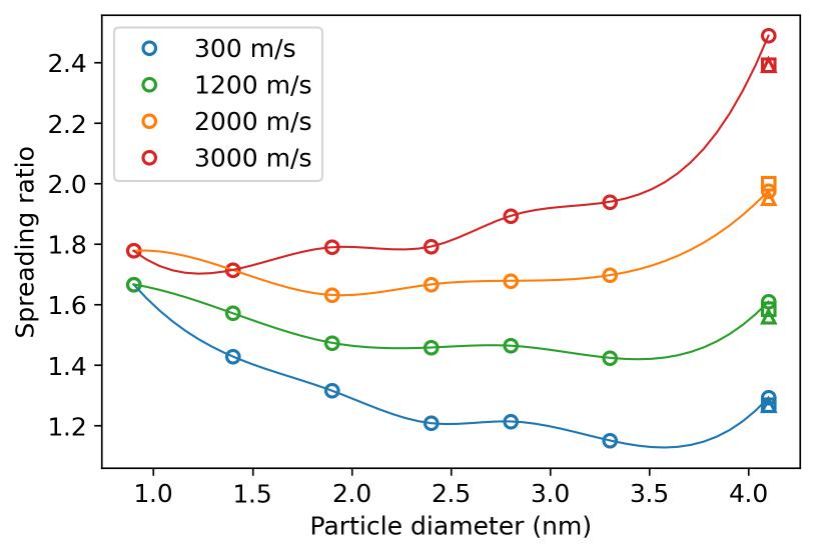}
    \caption{Variation of spreading ratio with the particle size and $V_{0}$. The lines are cubic splines to guide the eye. The circles, squares and triangles denote results on (100), (110), and (111)~Si substrates, respectively. }
    \label{fig:spreadfit}
\end{figure}

Furthermore, we calculated the spreading ratio as shown in figure~\ref{fig:spreadfit}. It can be seen that the variation in the particle size is nearly monotonic except for the largest particle. The ratios consist of reducing, constant and increasing trends. A lower $V_{0}$ gives a nearly reducing ratio,  while higher ones give more weight to the constant and increasing trends. To clarify whether or not a higher spreading ratio for 4.1~nm diameter particle is an artifact, we compared the results for different substrate sizes. We observed a negligible difference in the ratio between 100$\times$100 and 150$\times$150~{\AA}$^2$ substrate sizes. Besides, we plotted the 4.1~nm results on the (110) and (111)~Si substrates using square and triangle symbols, respectively. Again the result conforms with that for the (100) Si substrate, indicated by circles. We also compared different substrate dimensions (not shown here) and obtained similar results. Thus, the step change in the spreading ratio must be associated with the crossing of a critical diameter ($D_{\rm cr}$) where adiabatic heating becomes dominating giving rise to plastic deformation. \citet{schmidt2006} defined $D_{\rm cr}$ as a diameter above which thermal diffusion is slow enough for adiabatic shear instability to appear at the particle surface. They introduced $D_{\rm cr}=36\kappa/C_{\rm p}\rho V_0$ with $\kappa$, $C_{\rm p}$ and $\rho$ being thermal conductivity, heat capacity, and density of the particle. For typical Ti deposition spray velocities, their model leads to $D_{\rm cr}=400$~nm. However, the $V_{0}$ term in the denominator enables reducing the $D_{\rm cr}$ using higher $V_{0}$ such as in this study.

\begin{figure}[hbt!]
    \centering
    \includegraphics[width=1\linewidth]{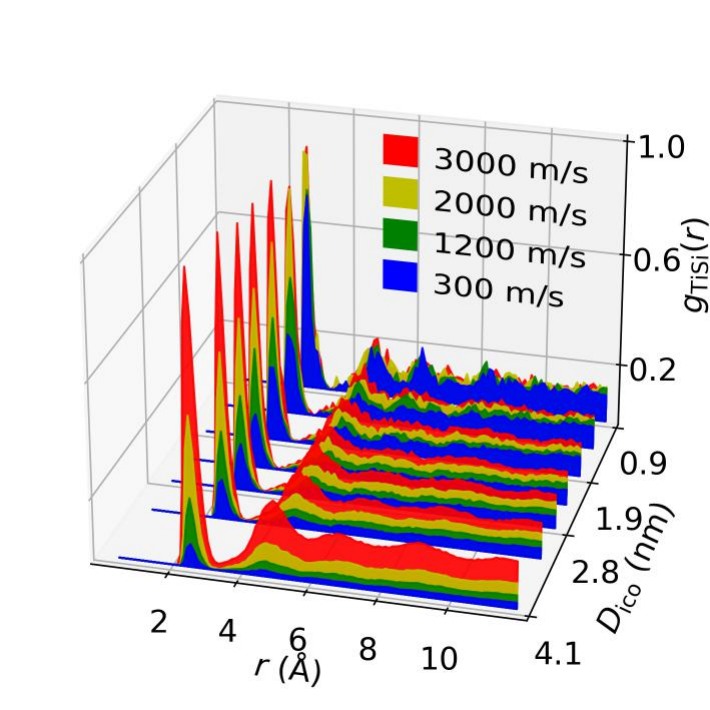}
    \caption{Partial radial distribution $g(r)$ for TiSi after spraying Ti ico particles of different diameters at different spray velocity $V_0$ onto the (100)~Si substrate.}
    \label{fig:gr_ico}
\end{figure}

Figure~\ref{fig:gr_ico} shows the variation of TiSi partial radial distribution $g(r)$ with ico particle size. It can be seen that for the smallest ico particle (0.9~nm) different $V_{0}$ does not change the intermixing. The difference between $g(r)$ due to variation in  $V_{0}$ is very small for the 1.4~nm diameter particle but its influence gradually increases with increased particle size until it becomes significant for the largest particle, here 4.1~nm in diameter. Thus, a higher $V_{0}$ becomes less important for the smaller particles. 

\begin{figure}[hbt!]
    \centering
    \includegraphics[width=1\linewidth]{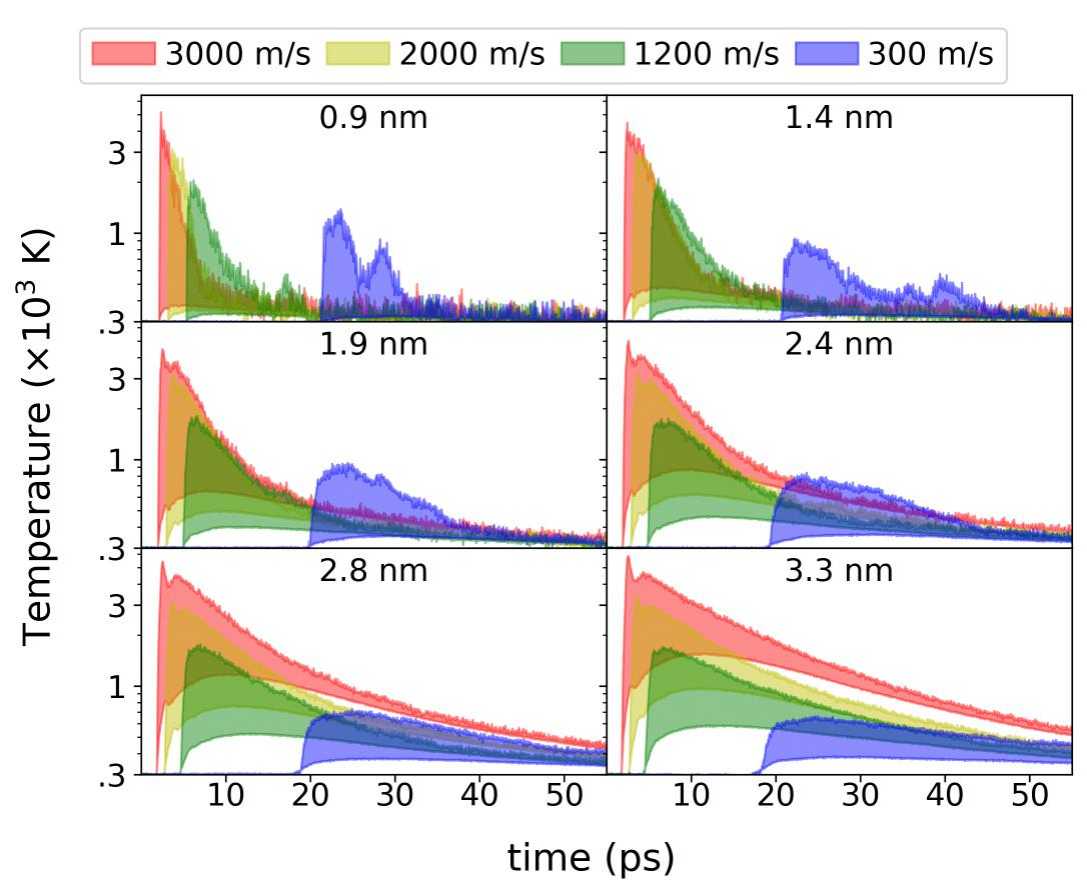}
    \caption{The temporal variation of the $T_{\rm par}$ (top limit) and $T_{\rm sur}$ (bottom limit) with varying spray velocity $V_{0}$ for Ti ico particles of various diameters onto a (100)~Si substrate. Note that the vertical axis is plotted in the semilog.}
    \label{fig:tempico}
\end{figure}

Figure~\ref{fig:tempico} shows the variation of $T_{\rm par}$ and $T_{\rm sur}$ with different spray velocities $V_{0}$ and particle sizes. Here the top and bottom limits denote $T_{\rm par}$ and $T_{\rm sur}$, respectively, and their difference is shaded. One can see immediately that $T_{\rm sur}^{\rm max}$ increases with increased particle size, while $T_{\rm par}^{\rm max}$ is independent of the particle size, and only dependent on the spray velocity $V_{0}$. Consider the temperature as $T=\frac{2}{3}k_{\rm B}\langle \mathcal{K}\rangle$ where the average kinetic energy is $\langle \mathcal{K}\rangle=\frac{1}{2}mV_0^2$ assuming adiabatic condition, at the early stage of impact. Thus, the particle's temperature is directly proportional to $V_0$ without being dependent on the number of atoms in the particle. The only exception is for spray velocity of  300~m/s as $T_{\rm par}^{\rm max}$ decreases with increased particle size. Earlier, \citet{hassani2019} reported increased temperature at the particle's perimeter, where the jet is expected, for larger particles. Note that here we plot the average particle's temperature. The atomistic temperature, cf.~figure~\ref{fig:tempshots}(a), indicates that during the flattening stage particle's perimeter has a higher temperature.

\begin{figure}[hbt!]
    \centering
    \includegraphics[width=1\linewidth]{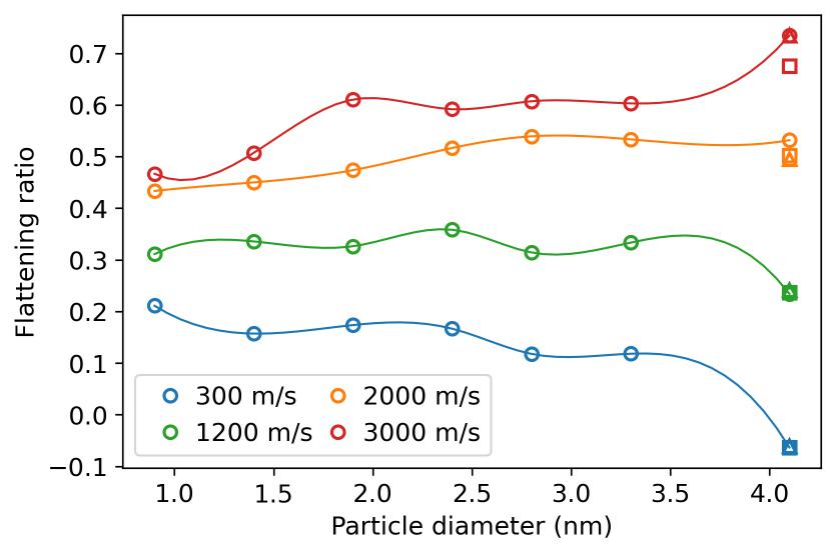}
    \caption{Variation of flattening ratio with the particle size and spray velocity $V_{0}$. The lines are cubic splines to guide the eye. The circles, squares and triangles denote results on (100), (110), and (111)~Si substrates, respectively. }
    \label{fig:flat}
\end{figure}

Figure~\ref{fig:flat} shows the flattening ratio, $1-h/D$, for different particle size and $V_{0}$. It can be seen that, unlike the spreading ratio, the step change at 4.1~nm is not always present. For adiabatic shear instability strains $\varepsilon>4$ \citep{assadi2003} and 4.5 \citep{hassani2018a} has been proposed using the Lagrangian and Eulerian finite element simulations. Comparing these numbers to the figure it appears that they refer to local values while the flatting ratio gives an average strain. For instance, the highest value here 0.7 corresponds to a final height of $\sim$1~nm. However, in-plane displacement of atoms can be much larger (cf.~figure~S3 of the supplementary materials).

\section{Conclusion}  \label{sec:summary}
We demonstrated cold spray of Ti nanoparticles (0.9 -- 4.1~nm in diameter) onto a Si substrate over a wide range of spray velocities $V_{0}$, using molecular dynamics simulations. By proper choice of thermostat layer and its damping, it is possible to model spray velocities $V_{0}$ up to 3000~m/s, which is necessary for the observation of adiabatic shear instability at the nanoscale. It is shown, that thanks to the ideally clean surfaces, bonding can be achieved over a wide range of spray velocities $V_{0}$ and particle sizes. However, for adiabatic shear instability to occur one needs at least a 4.1~nm diameter particle at 3000~m/s where a jet can be clearly observed. Here, we have been able to detect $D_{\rm cr}$ thanks to a step change in  the spreading ratio, $w/D$, while other strain measures, such as flattening, may not always show such a step change. Moreover, we demonstrated an oscillatory behavior of normal and radial stress in both space and time domains. The stress pulse duration and intensity change with the $V_{0}$ and depending on the $V_{0}$ may propagate across the interface. This behavior supports the recent hydrodynamic plasticity model, which serves as an alternative to adiabatic shear instability. Finally, the potential of cold spray to produce titanium silicide for electronic industry is demonstrated.

\section*{Supplementary Material}
The force field parameters compatible with LAMMPS, extra details on the calculation of splat diameter and supporting figures for different orientations of Si substrate and trajectory of atoms upon deformation are presented in supplementary materials. 

\url{https://doi.org/10.60893/figshare.jva.c.7496457}
\section*{AUTHOR DECLARATIONS}
The authors have no conflicts to disclose.
\section*{Data Availability}
The data that support the findings of this study are available from the corresponding author upon reasonable request.
\nolinenumbers
\bibliography{Ref}

\end{document}


\maketitle

\section{Tersoff parametrization}
We use Tersoff parameterization by \cite{plummer2019}. The parameter set compatible with LAMMPS is presented in Table~\ref{tab:tersoff}.

\begin{landscape}
\begin{table}[]
    \centering
\setlength{\tabcolsep}{3pt}
\begin{tabular}{c c c c c c c c c c c c c c c}
element & m & $\gamma$ & $\lambda_{3}$ & c & d & h & n & $\beta$ & $\lambda_{2}$ & B & R & D & $\lambda_{1}$ & A \\
Si Si Si & 1 & 0.092530 & 0.000000 & 1.136810 & 0.633970 & -0.335000 & 1 & 1 & 1.665910 & 361.557047 & 3.400000 & 0.150000 & 2.615479 & 1899.385778 \\
Si Si Ti & 1 & 0.059380 & 0.580472 & 0.647034 & 0.588344 & 1.061262 & 1 & 1 & 1.665910 & 361.557047 & 3.256477 & 0.297824 & 2.615479 & 1899.385778 \\
Si Ti Si & 1 & 0.092530 & 1.653831 & 1.136810 & 0.633970 & -0.335000 & 1 & 1 & 1.525709 & 230.770049 & 3.400000 & 0.150000 & 2.970786 & 3790.763031 \\
Si Ti Ti & 1 & 0.059380 & 0.602082 & 0.647034 & 0.588344 & 1.061262 & 1 & 1 & 1.525709 & 230.770049 & 3.256477 & 0.297824 & 2.970786 & 3790.763031 \\
Ti Si Si & 1 & 0.059380 & 1.653831 & 0.647034 & 0.588344 & 1.061262 & 1 & 1 & 1.525709 & 230.770049 & 3.256477 & 0.297824 & 2.970786 & 3790.763031 \\
Ti Si Ti & 1 & 0.001963 & 0.602082 & 1.356500 & 0.230100 & -0.904680 & 1 & 1 & 1.525709 & 230.770049 & 3.580900 & 0.302900 & 2.970786 & 3790.763031 \\
Ti Ti Si & 1 & 0.059380 & 0.145246 & 0.647034 & 0.588344 & 1.061262 & 1 & 1 & 1.367849 & 184.973776 & 3.256477 & 0.297824 & 1.940020 & 540.866546 \\
Ti Ti Ti & 1 & 0.001963 & 0.590960 & 1.356500 & 0.230100 & -0.904680 & 1 & 1 & 1.367849 & 184.973776 & 3.580900 & 0.302900 & 1.940020 & 540.866546 
    \end{tabular}
    \caption{Tersoff parametrization in the LAMMPS format (metal unit).}
    \label{tab:tersoff}
\end{table}
\end{landscape}

\section{Temperature}
\begin{figure}[hbt!]
    \centering
    \includegraphics[width=.7\linewidth]{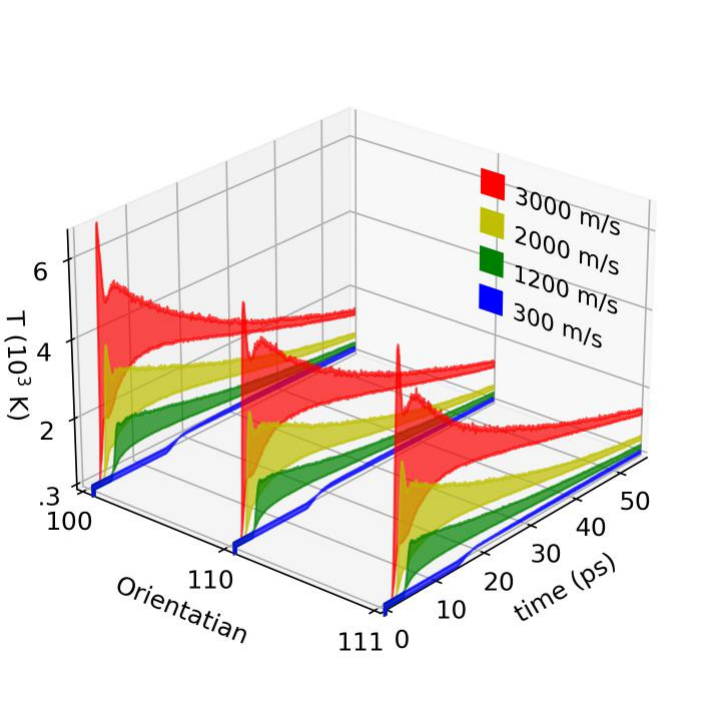}
    \caption{Variation of the particle and surface temperature with $V_{0}$ for 4.1~nm Ti particle on (100), (110) and (111)~Si substrates. In each shaded area top and bottom limits denote particle and surface layer temperatures, respectively.}
    \label{fig:tempico4.1}
\end{figure}

Figure~\ref{fig:tempico4.1} shows the temperature difference between $T_{\rm par}$ and $T_{\rm sur}$, i.e.\ upper and lower limits. One might predict that for higher planar density as in (111) the heat transfer towards the thermostat layer occurs more efficiently. However, at this range of temperature, this effect is negligible and we cannot see any difference. Thus we conclude that surface temperature remains the same for different Si orientations. 

\section{Spread}
\begin{figure}[hbt!]
    \centering
    \includegraphics[width=.6\linewidth]{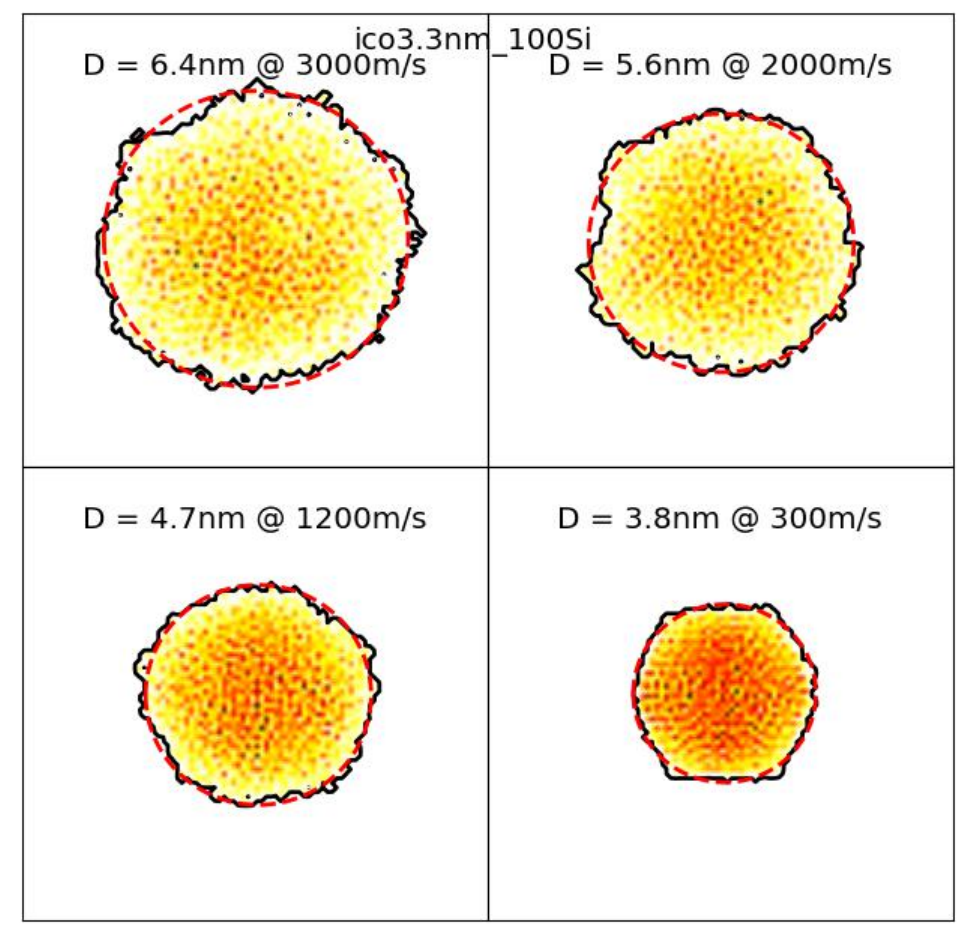}
    \caption{Illustration of circle fit (red dashed line) to the black contour of Ti atoms probability. The colorbar indicate tempo-spatial probability with its maximum in red and zero in white}
    \label{fig:fit}
\end{figure}

In order to determine the final diameter quantitatively, we produced a sum of Ti presence in 100-by-100 mesh which translates to the tempo-spatial probability of Ti. Then a circle fitted to a specific contour with a reasonable closed loop. The circle fit, contour and probability are shown in figure~\ref{fig:fit} for 3.3~nm Ti ico particle on (100)~Si.

\begin{figure}
    \centering
    \includegraphics[width=.7\linewidth]{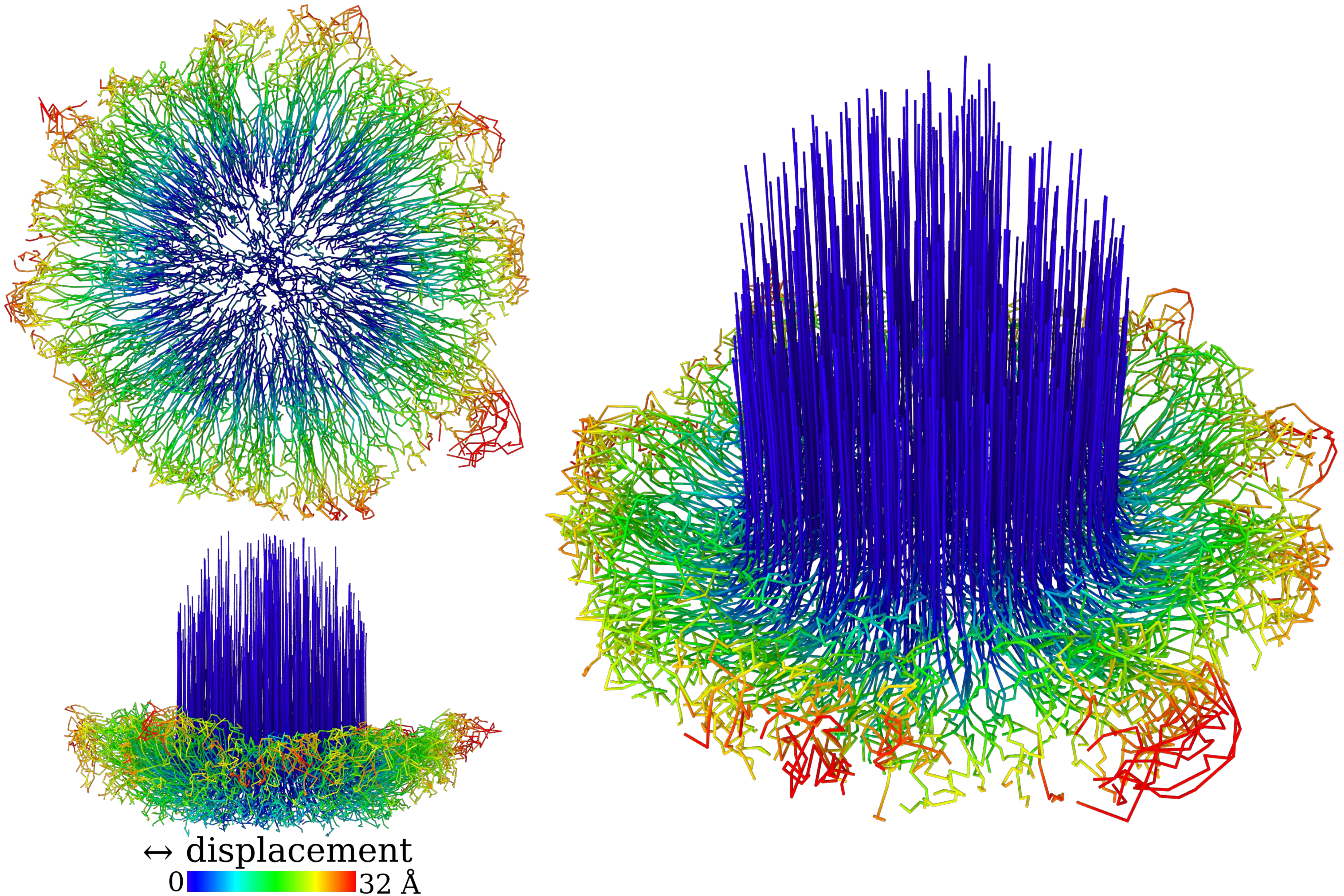}
    \caption{Top, side and tilt view of surface atoms trajectory lines for 4.1~nm ico cluster sprayed on (100)~Si substrate. The colorbar indicates the displacement magnitude in the plane.}
    \label{fig:flow}
\end{figure}

Figure~\ref{fig:flow} shows trajectory lines of surface atoms for 4.1~nm ico particle on the (100)~Si. 

\bibliographystyle{chicago}
\bibliography{Ref}